\documentclass[11pt,a4paper]{article}
\pdfoutput=1
\usepackage{jheppub}
\usepackage{amsfonts,amsbsy,amsmath,amssymb}
\usepackage{graphicx}
\usepackage{booktabs}
\usepackage{caption}
\usepackage{subcaption}
\usepackage{multirow}
\title{The gradient flow coupling in the Schr\"odinger Functional.}

\author[a]{Patrick Fritzsch}
\author[b]{and Alberto Ramos}

\affiliation[a]{Humboldt Universit\"at zu Berlin, Institut f\"ur Physik,
  Newtonstr. 15, 12489 Berlin, Germany.}
\affiliation[b]{NIC, DESY Platanenallee 6, 15738 Zeuthen, Germany.}
\emailAdd{fritzsch@physik.hu-berlin.de}
\emailAdd{alberto.ramos@desy.de}

\abstract{
We study the perturbative behavior of the Yang-Mills gradient flow in
the Schr\"odinger Functional, both in the continuum and on the lattice.
The energy density of the flow field is used to define a running coupling 
at a scale given by the size of the finite volume box. 
From our perturbative computation we estimate the size of 
cutoff effects of this coupling to leading order in perturbation theory. 
On a set of $N_{\rm f}=2$ gauge field
ensembles in a physical volume of $L\sim 0.4$~fm we finally
demonstrate the suitability of the  coupling for a precise  
continuum limit due to modest cutoff effects and high statistical precision.
}

\preprint{%
{\flushright 
DESY 12-241\\
HU-EP-12/53\\
SFB/CPP-13-05\\
}}

\begin{document}
\maketitle

\section{Introduction}

Finite-volume renormalization schemes have now a long history in
lattice field theory
(see~\cite{Luscher:1991wu,Luscher:1992an,deDivitiis:1994yp} or the
pedagogical reviews~\cite{Luscher:1998pe,Sommer:1997xw}). Asymptotic 
freedom tells us that at small distances QCD is well described by 
perturbation theory, while at large scales QCD is a strongly 
interacting theory. Instead of trying to accommodate these two scales 
in a single lattice simulation, the idea of finite-size scaling 
exploits the size of a finite volume world as renormalization scale. 
A single lattice simulation can resolve only a limited range of scales, but 
one can match different lattices and adopt a recursive procedure to cover a 
large range of scales. In this way one can connect the perturbative and 
non-perturbative regimes of QCD.

Beside a successful application of the finite-size scaling technique
in the case of pure Yang-Mills theory~\cite{Luscher:1992an, deDivitiis:1994yp}, 
the running coupling~\cite{Luscher:1992zx,Luscher:1993gh,DellaMorte:2004bc,Aoki:2009tf,Tekin:2010mm} 
and quark mass~\cite{Capitani:1998mq, DellaMorte:2005kg, Aoki:2010wm} have
been computed non-perturbatively in QCD with different flavour content 
by ALPHA and other collaborations. 
Since the general idea of finite-size scaling is a very powerful tool to 
solve scale dependent renormalization problems, it is not surprising that 
it is broadly used also in other strongly interacting theories, even in 
effective theories such as HQET~\cite{Heitger:2003nj, Sommer:2006sj, Sommer:2010ic}. 
There has been a growing interest in other than QCD strongly interacting 
gauge theories, especially in connection with electroweak symmetry breaking 
and quasi-conformal behavior (see for example~\cite{Neil:2012cb} and
references therein). Finite-size scaling techniques are also a powerful 
tool to study these systems. 

Basically there are two things that are needed to perform the previously 
sketched program. First one needs to define exactly what is meant by a 
finite-volume scheme, i.e., one has to specify the boundary conditions 
of the fields. Second, one needs a non-perturbative definition of the 
coupling. In principle there are many valid possibilities, but practical 
considerations have to be taken into account. Good options should allow 
for an easy evaluation of the coupling constant both in perturbation 
theory and in a numerical, non-perturbative (lattice) simulation.

The rest of this section is mainly dedicated to explain why we choose
the Schr\"odinger functional (SF) scheme~\cite{Luscher:1992an} as 
our finite-volume setup and the Wilson flow for a non-perturbative 
definition of the coupling~\cite{Luscher:2010iy}. To simplify the following 
discussion we will argue about a pure $SU(N)$ gauge theory in 4-dimensional 
Euclidean space-time. 

In the Schr\"odinger functional~\cite{Luscher:1992an,Sint:1993un}
one embeds the fields in a finite volume box of dimensions $L^3\times
T$. Gauge fields in the SF are periodic in the three spatial 
directions and have Dirichlet boundary conditions in time direction 
(i.e. one fixes the value of the gauge fields at $x_0=0, T$). The value
of the gauge fields at the time boundaries are called boundary
fields. One can interpret the partition function of the theory as the
transition amplitude of the gauge field to propagate from the
boundary value at $x_0=0$ to the boundary value at $x_0=T$. Such a 
setup has nice properties in perturbation theory. In particular with a 
smart choice of the boundary fields one can guarantee that there is a 
unique gauge field configuration (up to gauge transformations) that is 
a global minimum of the action. This avoids some difficulties with
perturbation 
theory~\cite{Coste:1985mn,GonzalezArroyo:1981vw,Luscher:1982ma}. 
The reader interested in this issue will appreciate the original
literature, as well as the nice discussion in~\cite{Fodor:2012qh}. 

Recently, the gradient flow has been used in different
contexts~\cite{Luscher:2009eq,Narayanan:2006rf,Lohmayer:2011si}, but 
it is the proposal made in~\cite{Luscher:2010iy} to define a renormalized
coupling through the gradient flow in non-abelian gauge theories what 
inspires this work. The gradient flow defines a family of gauge fields 
parametrized by a continuous flow time $t$. The flow equation brings the 
gauge field towards the minimum of the Yang-Mills action, and therefore 
represents a smoothing process. The key point is that correlation functions 
of the smoothed gauge field defined at $t>0$ are automatically 
finite~\cite{Luscher:2011bx}. One can use the expectation value of the
energy density, 
\begin{equation}
  \langle E(t) \rangle = \frac{1}{4}\langle G_{\mu\nu}(t)G_{\mu\nu}(t)\rangle \,,
\end{equation}
where $G_{\mu\nu}(t)$ is the field strength of the gauge field at
flow time $t$, to give a non-perturbative definition of the gauge
coupling. This idea was applied to set the scale in lattice
simulations~\cite{Luscher:2010iy,Borsanyi:2012zs}, to tune anisotropic
lattices~\cite{Borsanyi:2012zr} and more recently
in a similar context of this work (finite-size scaling, but using a
box with periodic boundary conditions) to compute the step scaling
function in $SU(3)$ with four fermion species~\cite{Fodor:2012td}. 

In this paper we investigate the perturbative behavior of the Wilson
flow in the Schr{\"o}dinger functional. This motivates us to propose a
gradient flow coupling
\begin{equation}
  \overline g_{\rm GF}^2(L) = \mathcal N^{-1} t^{2}\langle E(t)
  \rangle = \overline g_{\rm MS}^2 + 
  \mathcal O(\overline g_{\rm MS}^4),
\end{equation}
with a normalization factor $\mathcal N$ to be determined later, valid for
an arbitrary $SU(N)$ gauge field coupled (or not) to fermions. 
Relating $t$ and $L$ the
coupling depends only on one scale, the size of the finite volume box,
and therefore can be used for a finite-size scaling procedure in the same 
way as the traditional SF coupling. 

The paper is organized as follows: in the next section we investigate 
the perturbative behavior of $\langle E(t) \rangle$ in the SF, both in 
the continuum and on the lattice. Section~\ref{sc:coupling} uses this 
information to define the gradient flow coupling in the SF, and to 
discuss some practical issues: cutoff effects, boundary fields and 
fermions. In section~\ref{sc:nptests} we investigate this coupling 
numerically on a set of lattices in a physical volume of $L\sim 0.4$~fm 
and finally conclude in section~\ref{sc:conclusions}. Details needed for 
the computation have been summarized in form of appendices: a summary with 
some useful notation~\ref{ap:not}, heat kernels~\ref{ap:heat}, propagators 
in the SF~\ref{ap:prop} and finally some practical details on how to 
integrate the Wilson flow in numerical simulations~\ref{ap:int}.


\section{Perturbative behavior of the Wilson flow in 
the SF} 
\label{sc:pert}

We would like to start this section by recalling the original
proposal of using the Wilson flow and the energy density as a
definition for a coupling in gauge theories~\cite{Luscher:2010iy}. 
Later it will become clear what role the SF setup plays. 

\subsection{Generalities}

By considering the gauge fields to be functions of an extra flow
time $t$, not to be confused with Euclidean time, denoted
$x_0$, the Wilson flow is defined by the non-linear equation
\begin{equation}
  \label{eq:flow}
  \frac{ {\rm d}B_\mu(x,t)}{{\rm d}t} = D_\nu G_{\nu\mu}(x,t)\,, \qquad
  B_\mu(x,0) = A_\mu(x) \,,
\end{equation}
where 
\begin{eqnarray}
    G_{\mu\nu} &=& \partial_\mu B_\nu - \partial_\nu B_\mu + 
  [B_\mu,B_\nu] \,
\end{eqnarray}
is the field strength.
Due to $D_\nu G_{\nu\mu}\sim -\frac{\delta S_{\rm YM}[B]}{\delta B_\mu}$ 
gauge fields along the flow become smoother, eventually reaching a 
local minimum of the Yang Mills action: the flow smooths the fields 
over a region of radius $\sqrt{8t}$. 
The somewhat surprising result of~\cite{Luscher:2010iy,Luscher:2011bx}
is that correlation functions made of this smoothed field have a
well-defined continuum limit. In particular the energy density in
$SU(N)$ Yang-Mills theory in infinite volume has the perturbative
behavior 
\begin{equation}
  \langle E(t)\rangle = \frac{1}{4}\langle G_{\mu\nu}G_{\mu\nu}\rangle
  = \frac{3(N^2-1)\overline g_{\rm MS}^2}{128\pi^2t^2}(1+c_1\overline g_{\rm
    MS}^2+\mathcal O(\overline g_{\rm MS}^4))  \,.
\end{equation}
At a scale $\mu = 1/\sqrt{8t}$, $c_1$ is a numerical constant and
$\overline g_{\rm MS}(\mu)$ is the renormalized  
coupling in the ${{\rm MS}}$ scheme. Therefore one can define a running 
coupling constant $\alpha(\mu)$ from
\begin{equation}
  t^2\langle E(t)\rangle =
  \frac{3(N^2-1)}{32\pi}\alpha(\mu) \,.
\end{equation}
These expressions are valid in infinite volume. What about the 
Schr{\"o}dinger Functional? The computation is completely analogous, but we
have to impose the correct boundary conditions to the gauge fields. As
we have mentioned in the SF gauge fields are restricted to a box
of dimensions $L^3\times T$. They are periodic in the three spatial 
directions and the spatial components have Dirichlet boundary conditions
at $x_0=0$ and $x_0=T$. We are going to work exclusively with zero
boundary fields, which means 
\begin{eqnarray}
  \label{eq:bc1}
  B_\mu(x+\hat kL, t) &=&   B_\mu(x,t)    \,,\\
  \label{eq:bc2}
  B_k(x, t)|_{x_0=0, T} &=&  0            \,.
\end{eqnarray}
The flow equation~\eqref{eq:flow} has to be solved maintaining these
boundary conditions at all flow times $t$. To apply the idea of
finite-size scaling,  
as has previously been done in~\cite{Fodor:2012qh} in a periodic box,
one simply has to run the renormalization scale with the size of  
the finite volume box given by $L$ via
\begin{equation} \label{eq:scale}
  \mu = \frac{1}{\sqrt{8t}} = \frac{1}{cL} \,.
\end{equation}
Here $c$ is a dimensionless constant that represents the fraction of
the smoothing range over the total size of the box. In this way the
flow coupling will not depend on any scale other than $L$. 
The renormalization scheme will depend on the values of $c$, $\rho=T/L$ and%
\footnote{Note that in the SF the boundary conditions break the invariance 
under time translations. Therefore $\langle E(t,x_0)\rangle$ will depend 
explicitly on $x_0$.} $x_0/T$ 
\begin{equation}
  \overline g^2_{\rm GF}(L) = \mathcal N^{-1}(c,\rho,x_0/T) t^2\langle
  E(t,x_0)\rangle\Big|_{t=c^2L^2/8 }  \,,
\end{equation}
where $\mathcal N^{-1}(c,\rho,x_0/T)$ will be computed in the next section 
in order to ensure
\begin{equation}
  \overline g^2_{\rm GF} = g_0^2 + \mathcal O(g_0^4)  \,.  
\end{equation}

\subsection{Continuum}

Our computation follows the lines of~\cite{Luscher:2011bx}. First we
consider the modified flow equation
\begin{equation}
  \label{eq:flowgauge}
  \frac{{\rm d} B_\mu}{{\rm d}t} = D_\nu G_{\nu\mu} + 
  \alpha D_\mu\partial_\nu B_\nu \,, \qquad
  B_\mu(x,0) = A_\mu(x) \,.
\end{equation}
One can transform a solution of the last equation into a solution of
the canonical flow equation~\eqref{eq:flow} (corresponding to
$\alpha=0$) by a flow-time dependent gauge transformation. 
In particular, if $B_\mu$ is a solution of~\eqref{eq:flowgauge}
one can construct a solution of~\eqref{eq:flow} via
\begin{equation}
  B_\mu\big|_{\alpha=0} = \Lambda B_\mu\Lambda^{-1} + 
  \Lambda \partial_\mu
  \Lambda^{-1} 
\end{equation}
as long as $\Lambda$ obeys the equation
\begin{equation}
  \frac{{\rm d} \Lambda}{{\rm d}t} =
  \alpha \Lambda \partial_\mu B_\mu \,;\quad
  \Lambda\big|_{t=0} = 1\,.
\end{equation}
This shows that gauge-invariant quantities are independent of
$\alpha$. For instance, setting $\alpha=1$ turns out to be a 
very convenient choice for perturbative computations. 
Due to the periodicity in the spatial directions it is natural to
expand the gauge fields as
\begin{equation}
  A_\mu(x) = \frac{1}{L^3}\sum_{\mathbf p} e^{\imath \mathbf
    p\cdot\mathbf x}\tilde A_{\mu}(\mathbf p,x_0) \,.
\end{equation}
As already mentioned, in the SF the gauge field is periodic in the 
three spatial directions and its spatial components have Dirichlet 
boundary conditions in time, eq.~\eqref{eq:bc1} and~\eqref{eq:bc2} 
respectively.
On the other hand the boundary conditions of the time component of the
gauge field are not fixed but naturally emerge through the gauge
fixing condition.\footnote{The authors want to thank M. L\"uscher for
helping us to understand this point.} To properly derive the
boundary conditions for $B_0$ it is convenient to work in the lattice
formulation and derive the boundary conditions by taking the
continuum limit. We will postpone this derivation to the next section
and simply state the result here: 
$B_0$ obeys Neumann boundary conditions at non-vanishing spatial momentum, 
while for zero momentum $B_0$ obeys mixed boundary conditions. Thus
in the present set-up the full set of boundary conditions reads 
\begin{subequations}
  \label{eq:bc}
\begin{align}
  \forall\mathbf{p}: &&             \tilde B_k(\mathbf{p}, x_0, t)|_{x_0=0,T}            &=0   \;, \\
  \mathbf{p} \ne 0 : &&  \partial_0 \tilde B_0(\mathbf{p}, x_0, t)|_{x_0=0,T}            &=0   \;, \\
  \mathbf{p}   = 0 : &&             \tilde B_0(\mathbf{0}, x_0, t)|_{x_0=0\hphantom{,T}} &=0   \;,
                      &  \partial_0 \tilde B_0(\mathbf{0}, x_0, t)|_{x_0=T}              &=0   \;. \label{eq:cont-B0-p=0}
\end{align}
\end{subequations}

The modified Wilson flow equation with $\alpha=1$ is given by
\begin{equation}
  \frac{{\rm d} B_\mu}{{\rm d}t} = D_\nu G_{\nu\mu} +
  D_\mu\partial_\nu B_\nu \,.
\end{equation}
After rescaling the gauge potential with the bare coupling $A_\mu
\rightarrow g_0A_\mu$, the flow becomes a function of the coupling
\begin{equation}
  \tilde B_\mu(\mathbf p, x_0, t) = 
  \sum_{n=1}^{\infty} \tilde B_{\mu,n}(\mathbf p, x_0, t) g_0^n \,.
\end{equation}
Inserting this expression in the modified flow equation, we find that
to leading order in $g_0$ the flow equation is just the heat equation 
with initial condition $A_{\mu}$:
\begin{eqnarray}
   \frac{{\rm d} \tilde B_{\mu,1}(\mathbf p, x_0, t)}{{\rm d}t} &=&
   (-\mathbf p^2 + \partial_0^2) 
   \tilde B_{\mu,1}(\mathbf p, x_0, t) \\
   \tilde B_{\mu,1}(\mathbf p, x_0, 0) &=& \tilde A_{\mu}(\mathbf p, x_0)\, ,
\end{eqnarray}
i.e., to leading order the Wilson flow is the heat flow. We also
observe that different momentum modes do not couple to each other 
at this order.
Together with the fact that the zero
momentum mode $B_0(\mathbf{0}, x_0, t)$ does not contribute to the
observable of interest, $E(t)=\frac{1}{4}G_{\mu\nu}G_{\mu\nu}$, we can
safely neglect the special treatment that the boundary conditions of
the zero momentum mode
$B_0(\mathbf{0}, x_0, t)$ would otherwise require in the following discussion.

We have to
solve the heat equation respecting the boundary conditions~\eqref{eq:bc}.
This is easily done by using appropriate heat kernels
\begin{subequations}
  \label{eq:btexp}
  \begin{eqnarray}
    \tilde B_{k,1}(\mathbf p, x_0, t) &=& e^{-\mathbf p^2 t}
    \int_0^T \!\!{\rm d}x'_0\, 
    K^{D}(x_0,x_0',t) \tilde A_{k}(\mathbf p, x'_0) \,,\\ 
    \tilde B_{0,1}(\mathbf p, x_0, t) &=& e^{-\mathbf p^2 t}
    \int_0^T \!\!{\rm d}x'_0\, 
    K^{N}(x_0,x_0',t) \tilde A_{0}(\mathbf p, x'_0)\quad (\mathbf p\ne
    \mathbf 0) \,. 
  \end{eqnarray}
\end{subequations}
Since the boundary conditions of the field $\tilde B_{\mu,1}(\mathbf p, x_0, t)$
are inherited from the boundary conditions of the heat kernels, we
have to choose them with the correct boundary conditions. Heat kernels
with either Dirichlet ($K^D(x,x',t)$) or Neumann ($K^N(x,x',t)$)
boundary conditions can be constructed from the basic periodic
($K^P(x,x',t)$) heat kernel in $[0,L]$ given by 
\begin{equation}
  K^P(x,x',t) = \frac{1}{L}\sum_p e^{-p^2t}e^{\imath p(x-x')}, \qquad
  \left(p=\frac{2\pi n}{L};\, n\in\mathbb Z\right)\,.
\end{equation}
Explicit expressions are given in appendix~\ref{ap:heat}. 

Our observable, the energy density $\langle E(t,x_0) \rangle$, has an 
expansion in powers of $g_0$. The leading contribution is given by 
\begin{equation}
  \label{eq:e0}
  \mathcal E_0(t,x_0) = \frac{g_0^2}{2}\langle
  \partial_{\mu}B_{\nu,1}^a\partial_{\mu}B_{\nu,1}^a - 
  \partial_{\mu}B_{\nu,1}^a\partial_{\nu}B_{\mu,1}^a
  \rangle \,.
\end{equation}
We are going to split the computation in two parts, one involving only
the spatial components of $G_{\mu\nu}$, and the other involving the
mixed time-space components of $G_{\mu\nu}$
\begin{eqnarray}  \label{eq:e0s}
  \mathcal E_0^s(t,x_0) &=& \frac{g_0^2}{2}\langle
  \partial_{i}B_{k,1}^a\partial_{i}B_{k,1}^a - 
  \partial_{i}B_{k,1}^a\partial_{k}B_{i,1}^a
  \rangle \,,\\ \label{eq:e0m}
  \mathcal E_0^m(t,x_0) &=& \frac{g_0^2}{2}\langle
  \partial_{0}B_{k,1}^a\partial_{0}B_{k,1}^a - 
  \partial_{0}B_{k,1}^a\partial_{k}B_{0,1}^a
  \rangle \,.
\end{eqnarray}
Inserting for instance expression~\eqref{eq:btexp} into~\eqref{eq:e0s} we obtain
\begin{eqnarray}
  \nonumber
  \mathcal E_0^s(t,x_0) &=&  -\frac{g_0^2}{2L^6}
  \sum_{\mathbf
    p ,\mathbf q}e^{-t(\mathbf p^2+\mathbf q^2)}
  e^{\imath (\mathbf p+\mathbf q)\mathbf x} \!\!\int_0^T {\rm
    d}x_0' {\rm d}y_0' 
  \,  K^D(x_0,x_0',t) K^D(x_0,y_0',t)  \\ 
  && \times \left[ 
  p_iq_i \langle \tilde A_k^a(\mathbf p, x_0')\tilde A_k^a(\mathbf
  q,y_0')\rangle - p_iq_k \langle \tilde A_i^a(\mathbf p, x_0')\tilde
  A_k^a(\mathbf q,y_0')\rangle
  \right]  \,.
\end{eqnarray}
The final result is obtained inserting the SF gluon 
propagator~\cite{Luscher:1996vw,Weisz:int1996}. Since our observable
is invariant under gauge transformations of the $A_\mu(x)$ field
we will use the Feynman gauge, where the 
expression for the gluon propagator turns out to be more easy (for
additional details see appendix~\ref{ap:prop})\footnote{We have
  checked that the result is independent of the gauge choice.}. 
\begin{equation}
  \langle\tilde A_i^a(\mathbf p,x_0) \tilde A_k^b(\mathbf q,y_0)
  \rangle  = L^3\delta_{ab}\delta_{ik}\delta_{\mathbf p,- \mathbf q} 
    \frac{1}{T}\sum_{p_0}
  \frac{s_{p_0}(x_0)s_{p_0}(y_0)}{{\mathbf p}^2 +
    \left(\frac{p_0}{2}\right)^2}  +\mathcal O(g_0^2)\,.
\end{equation}
To shorten notation we use
\begin{align}
  s_{p_0}(x) &= \sin\left(\frac{p_0 x}{2}\right)  \,, &  
  c_{p_0}(x) &= \cos\left(\frac{p_0 x}{2}\right)  \,, &
  p_0&=\frac{2\pi n_0}{T} \,.
\end{align}
After some algebraic work one arrives at the expression
\begin{eqnarray}
    t^2\mathcal E_0^s(t,x_0)\Big|_{t=c^2L^2/8 } = 
  \frac{c^4(N^2-1)g_0^2}{64\rho} \sum_{\mathbf n, n_0} 
  e^{-c^2\pi^2 (\mathbf n^2 + \frac{1}{4\rho^2} n_0^2) }
  \frac{\mathbf n^2  }{\mathbf n^2 + \frac{1}{4\rho^2}n_0^2} s_{n_0}^2(x_0)  
\end{eqnarray}
and a very similar computation leads to
\begin{eqnarray}
    t^2\mathcal E_0^m(t,x_0)\Big|_{t=c^2L^2/8 } = 
  \frac{c^4(N^2-1)g_0^2}{128\rho} \sum_{\mathbf n, n_0} 
  e^{-c^2\pi^2 (\mathbf n^2 + \frac{1}{4\rho^2} n_0^2) }
  \frac{\mathbf n^2 + \frac{3}{4\rho^2}n_0^2}{
        \mathbf n^2 + \frac{1}{4\rho^2}n_0^2} c_{n_0}^2(x_0)  \,.
\end{eqnarray}

\subsection{Lattice}

On the lattice one defines the Wilson flow as
\begin{equation}
  \label{eq:flowlat}
  a^2\partial_t V_\mu(x,t) = -g_0^2 \{T^a\partial_{x,\mu}^a S_w(V)\}
  V_\mu(x,t) \,,  \qquad V_\mu(x,0) = U_\mu(x)  \,.
\end{equation}
If
$f(U_\mu(x))$ is an arbitrary function of the link variable
$U_\mu(x)$, the components of its Lie-algebra valued derivative
$\partial_{x,\mu}^a $  
are defined as 
\begin{equation}
   \partial_{x,\mu}^a f(U_\mu(x)) = \left.\frac{ {\rm d} f(e^{\epsilon
            T^a}U_\mu(x))}{ {\rm d}\epsilon} \right|_{\epsilon=0}\,. 
\end{equation}
In a neighborhood of the classical vacuum configuration the lattice 
fields $U_\mu(x)$ and $V_\mu(x,t)$ are parametrized as follows:
\begin{align}
  U_\mu(x)   &= \exp\{ag_0 A_\mu(x)\}   \;, &
  V_\mu(x,t) &= \exp\{ag_0 B_\mu(x,t)\} \;.
\end{align}

\subsubsection{Gauge fixing}

To simplify our perturbative computations it is useful to
study a modified equation with a gauge damping term. It is easy to
check that the lattice flow equation~\eqref{eq:flowlat} is invariant
under flow-time independent gauge transformations. On the other hand
one can consider the modified equation   
\begin{equation}
  \label{eq:flowlatmd}
  a^2\partial_t V_\mu^\Lambda(x,t) = g_0^2 \left\{ 
        -\big[ T^a\partial_{x,\mu}^a S_w(V^\Lambda) \big] 
        + a^2\hat D_\mu^{\Lambda}\big[\Lambda^{-1}(x,t)\dot \Lambda(x,t)\big] 
                                           \right\} V_\mu^\Lambda(x,t) \,,
\end{equation}
with $V_\mu^\Lambda(x,0) = U_\mu(x)$ and the forward lattice covariant
derivative 
$\hat D_\mu^{\Lambda}$ acting on Lie-algebra valued functions according to
\begin{equation}
  \hat D_\mu f(x) = \frac{1}{a}\left[
    V_\mu(x,t)f(x+\hat\mu)V_\mu^{-1}(x,t) - f(x)
  \right] \;. 
\end{equation}
With $\hat \partial, \hat \partial^*$ we denote the forward/backward 
finite differences respectively as defined in appendix~\ref{ap:not}.

The solutions of the modified equation~\eqref{eq:flowlatmd} and the
original flow equation~\eqref{eq:flowlat} are related by a gauge 
transformation 
\begin{equation}
  V_\mu(x,t) = \Lambda(x,t)V_\mu^\Lambda(x,t)\Lambda^{-1} (x+\hat\mu,t)
\end{equation}
and therefore one can freely choose the function $\Lambda(x,t)$. To
fix the gauge the most natural choice is to use the same 
functional that is used for the conventional gauge fixing. As is
detailed in appendix~\ref{ap:prop}, we choose 
\begin{equation}
  \label{eq:lam}
  \Lambda^{-1}\frac{{\rm d} \Lambda}{{\rm d}t} = \left\{
    \begin{array}{ll}
            \alpha \hat\partial^*_\mu B_\mu(x,t)   & \text{ if } 0<x_0<T \,,\\
            {\alpha}\frac{a^2}{L^3}\sum_{\mathbf x}B_0(x,t) &
            \text{ if } x_0=0 \\ 
            0                                     & \text{ if } x_0=T \\
    \end{array}
  \right.
\end{equation}
with initial condition
\begin{equation}
  \Lambda\big|_{t=0} = 1\,.
\end{equation}
Note $\Lambda(x,t)$ does not depend on $\mathbf x$ at $x_0=0$ and
$x_0=T$, as a decent gauge transformation should be in the 
Schr\"odinger functional according to our conventions (see
appendix~\ref{ap:prop} for details).  

We observe that on the lattice the time component of the gauge field
$B_0(x,t)$ is completely free and does not obey any particular
boundary conditions. To understand how the boundary conditions for
$B_0(x,t)$ arise in the continuum theory, one can extend the
domain of definition of $B_0(x,t)$ to $-a\le x_0 \le T$ and choose to
fix the additional variables with the condition
\begin{eqnarray}
  \hat \partial_0^* B_0(x,t) = \left\{
    \begin{array}{ll}
            \frac{a^2}{L^3}\sum_{\mathbf x}B_0(x,t) & \text{ if } x_0=0  \,,\\
                                                0 & \text{ if } x_0=T  \,.
    \end{array}
    \right.
\end{eqnarray}
This equation can be interpreted as a boundary condition for the
$B_0(x,t)$ field. 
In particular $B_0(x,t)$ has Neumann boundary
conditions at $x_0=0,T$, except for its spatial momentum zero mode 
that has a mixture of Neumann boundary conditions at $x_0=T$ and 
Dirichlet boundary conditions at $x_0=-a$. 
\begin{align}
  \mathbf{p}\ne 0 : &&  \hat \partial_0^*\tilde B_0(\mathbf p, x_0, t)|_{x_0=0,T} &=0  \,, \\  \nonumber
  \mathbf{p}  = 0 : &&                   \tilde B_0(\mathbf 0, x_0, t)|_{x_0=-a}\,&=0  \,, 
                     &  \hat \partial_0^*\tilde B_0(\mathbf 0, x_0, t)|_{x_0=T}   &=0  \,.
\end{align}
This justifies our previous choice of boundary conditions in the
continuum, eq.~\eqref{eq:bc}. With this useful convention in mind
eq.~\eqref{eq:lam} simply reads 
\begin{equation}
  \label{eq:lam2}
  \Lambda^{-1}\frac{{\rm d} \Lambda}{{\rm d}t} = \alpha
  \hat\partial^*_\mu B_\mu(x,t)\,.
\end{equation}

\subsubsection{Behaviour of $\langle E(t)\rangle$ in lattice perturbation theory}

We again note that the value of any gauge invariant observable is
independent of our choice of $\alpha$ in equation~\eqref{eq:lam2}. In
particular, with the choice $\alpha=1$ the modified flow equation
reads 
\begin{equation}
  a^2\partial_t V_\mu(x,t) = g_0^2 \left\{ -[T^a\partial_{x,\mu}^a
      S_w(V)] + a^2\hat D_\mu(\hat\partial_\nu^* B_\nu ) 
      \right\}
  V_\mu(x,t) \,,  \qquad V_\mu(x,0) = U_\mu(x) \,,
\end{equation}
and to first order in $g_0$
\begin{equation}
  \label{eq:wflowlato1}
  \partial_t B_{\mu,1}(x,t) = \hat \partial_\nu\hat\partial_\nu^*
  B_{\mu,1}(x,t) \,.
\end{equation}
Using periodicity in the spatial directions, we expand 
\begin{eqnarray}
  B_i(x,t) &=& \frac{1}{L^3}\sum_{\mathbf p} e^{\imath \mathbf
    p\cdot\mathbf x}e^{\imath a p_i/2}\tilde B_{i}(\mathbf p,x_0,t) \,,\\
  B_0(x,t) &=& \frac{1}{L^3}\sum_{\mathbf p} e^{\imath \mathbf
    p\cdot\mathbf x}\tilde B_{0}(\mathbf p,x_0,t)                   \,,
\end{eqnarray}
and the flow equation becomes
\begin{equation}
  \partial_t \tilde B_{\mu,1}(\mathbf p, x_0,t) =
  (-\mathbf{\hat p}^2  + \hat \partial_0\hat\partial_0^*) 
  \tilde B_{\mu,1}(\mathbf p, x_0, t) \,,
\end{equation}
where $\mathbf{\hat p}$ is the usual spatial lattice momentum, see
appendix~\ref{ap:not}.

Now we have to solve a 
special type of heat equation in which the Laplacian is
substituted by a discrete version, but the flow time remains
a continuous variable. The strategy is very similar: We find the
fundamental solutions of this equation, i.e., the discrete heat
kernels given in appendix~\ref{ap:heat}, and write
\begin{eqnarray}
      \tilde B_{k,1}(\mathbf p, x_0, t) &= &
      e^{-\hat {\mathbf p}^2 t}\sum_{x'_0=0}^T \, 
      \hat K^{D}(x_0,x_0',t)
      \tilde A_{k}(\mathbf p, x'_0)  \,, \\
      \tilde B_{0,1}(\mathbf p, x_0, t) &= &
      e^{-\hat {\mathbf p}^2 t}\sum_{x'_0=0}^T \, 
      \hat K^{N}(x_0,x_0',t)
      \tilde A_{0}(\mathbf p, x'_0)\quad (\mathbf p\ne 0)  \,.
\end{eqnarray}
Then we have to insert this in our lattice observable $\langle
E\rangle$. We use the clover definition for $G_{\mu\nu}$ that to
leading order in $g_0$ reads
\begin{equation}
  G_{\mu\nu} = \frac{g_0}{2}\,\widetilde\partial_\mu\left[B_{\nu,1}(x) + 
  B_{\nu,1}(x-\hat \nu)\right] -
               \frac{g_0}{2}\,\widetilde\partial_\nu\left[B_{\mu,1}(x) + 
  B_{\mu,1}(x-\hat \mu)\right] + \mathcal O(g_0^2) \,,
\end{equation}
where $\widetilde\partial_{\mu} = \tfrac{1}{2}(\hat \partial_\mu +
\hat \partial^*_\mu)$. The computation is completed by using the
lattice gluon propagator 
\begin{equation}
  \langle\tilde A_i^a(\mathbf p,x_0) \tilde A_k^b(\mathbf q,y_0)
  \rangle  = L^3 \delta_{ab}\delta_{ik}\delta_{\mathbf p,- \mathbf q}\, 
    \frac{1}{T}\sum_{p_0}
  \frac{\hat s_{p_0}(x_0)\hat s_{p_0}(y_0)}{\hat{\mathbf p}^2 +
    {\check p_0}^2}  +\mathcal O(g_0^2)\,.
\end{equation}
For the spatial part of the contribution to the energy density we 
arrive at
\begin{eqnarray}
  \label{eq:lats}
  \nonumber
  t^2\hat{\mathcal E}_0^s(t,x_0)\Big|_{t=c^2L^2/8 } &=& 
  \frac{(N^2-1)c^4g_0^2}{128 \rho} \sum_{\mathbf p, p_0}
  e^{-\frac{L^2c^2}{4}(\hat{\mathbf p}^2 + \check p_0^2)} \times\\
&&  \frac{\mathring{\mathbf p}^2\cos^2(ap_i/2) - (\mathring
    p_i\cos(ap_i/2))^2 }{\hat{\mathbf p}^2 +
    \check p_0^2}\hat s_{p_0}^2(x_0)\,,
\end{eqnarray}
while for the mixed part we obtain
\begin{eqnarray}
  \label{eq:latm}
  \nonumber
  t^2\hat{\mathcal E}_0^m(t,x_0)\Big|_{t=c^2L^2/8 } &=& 
  \frac{(N^2-1)c^4g_0^2}{128 \rho} \sum_{\mathbf p, p_0}
  e^{-\frac{L^2c^2}{4}(\hat{\mathbf p}^2 + \check p_0^2)} \times\\
  && \frac{\mathring{\mathbf p}^2\cos^2(ap_0/4) + \frac{1}{4}\hat p_0^2\cos^2(ap_i/2) }{\hat{\mathbf p}^2 +
    \check p_0^2}\hat c_{p_0}^2(x_0-a/2)  \,.
\end{eqnarray}
The definitions of the lattice momenta $\hat p$, $\check p$, $\mathring p$ 
and the functions $\hat s_p(x), \hat c_p(x)$ are summarized in
appendix~\ref{ap:not}.

\subsection{Tests}

There are several tests that can be performed to check the previous
computations. Let us first concentrate on the continuum
computation. At fixed $t$, the infinite volume limit $L\rightarrow
\infty$ (with $\rho$ kept constant) is taken through $c\rightarrow 0$. 
For this case the continuum expression transforms into an integral via
\begin{eqnarray*}
  c\mathbf n \longrightarrow  \mathbf p \,,\qquad
  \frac{c}{\rho}k \longrightarrow  p_0  \,,\qquad
  \frac{c^4}{\rho}\sum_{\mathbf n, k} \longrightarrow  \int {\rm d}^4p \,,
\end{eqnarray*}
and we obtain
\begin{eqnarray}
  \nonumber
  \lim_{c\rightarrow 0} \left[t^2\mathcal E_0^s(t,T/2) +
  t^2\mathcal E_0^m(t,T/2)\right] &=&  
  \frac{g_0^2(N^2-1)}{128}\int d^4p\, e^{-\pi^2(\mathbf p^2 + p_0^2)}
  \frac{2\mathbf p^2 + \mathbf p^2 + 3p_0^2}{\mathbf p^2+p_0^2}\\ 
  &=&
  \frac{3g_0^2(N^2-1)}{128\pi^2}
\end{eqnarray}
thus recovering the infinite volume result of~\cite{Luscher:2011bx}. 

Another rather obvious check is that one should recover the continuum
result from the lattice expression in the limit $a/L\rightarrow 0$. This
can be easily checked by noting that the sums~\eqref{eq:lats} and
~\eqref{eq:latm} are dominated by terms with small lattice momenta
$ap_\mu\rightarrow 0$. 

\begin{figure}
  \centering
    \begin{subfigure}[b]{0.49\textwidth}
    \includegraphics[width=\textwidth]{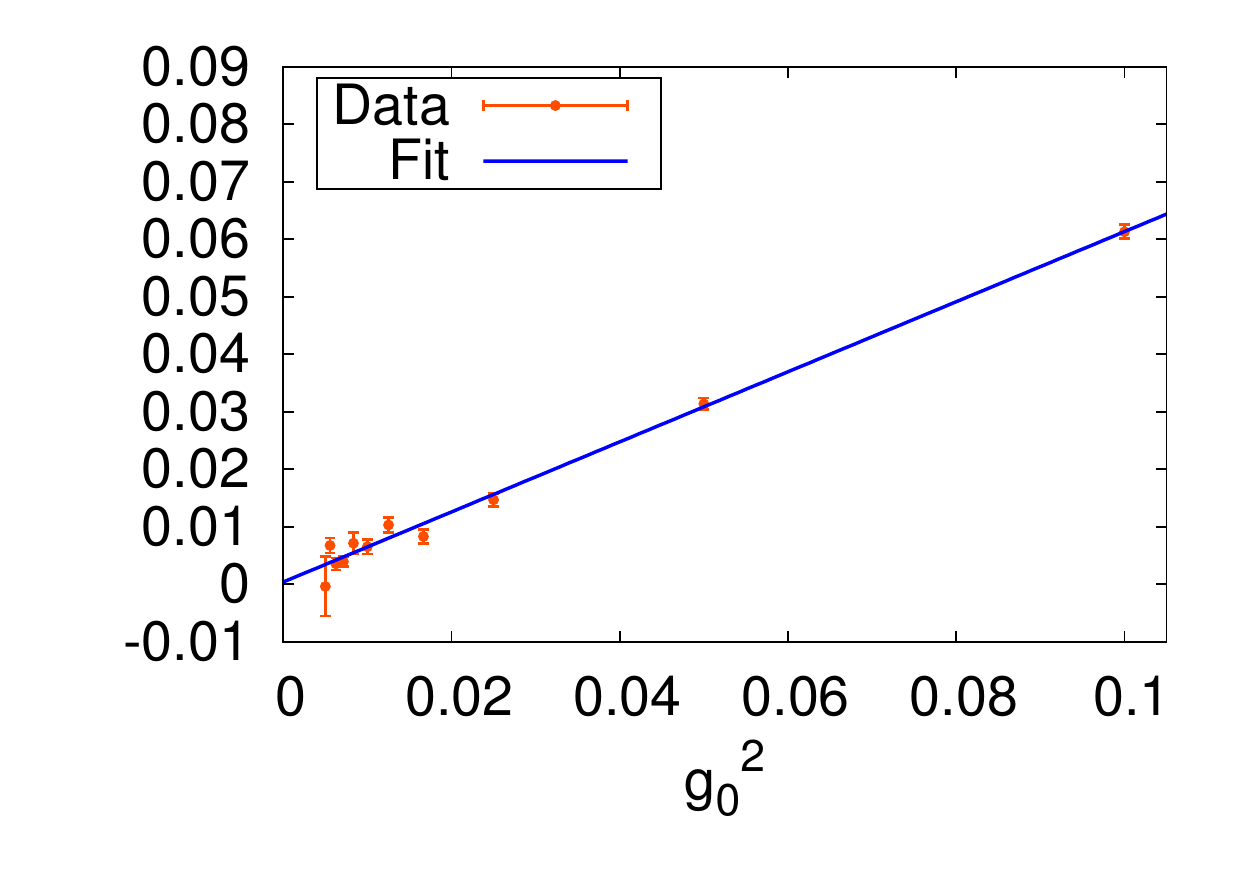}
    \caption{$c=0.3$ and $x_0/a=6$.}
  \end{subfigure}
  \begin{subfigure}[b]{0.49\textwidth}
    \includegraphics[width=\textwidth]{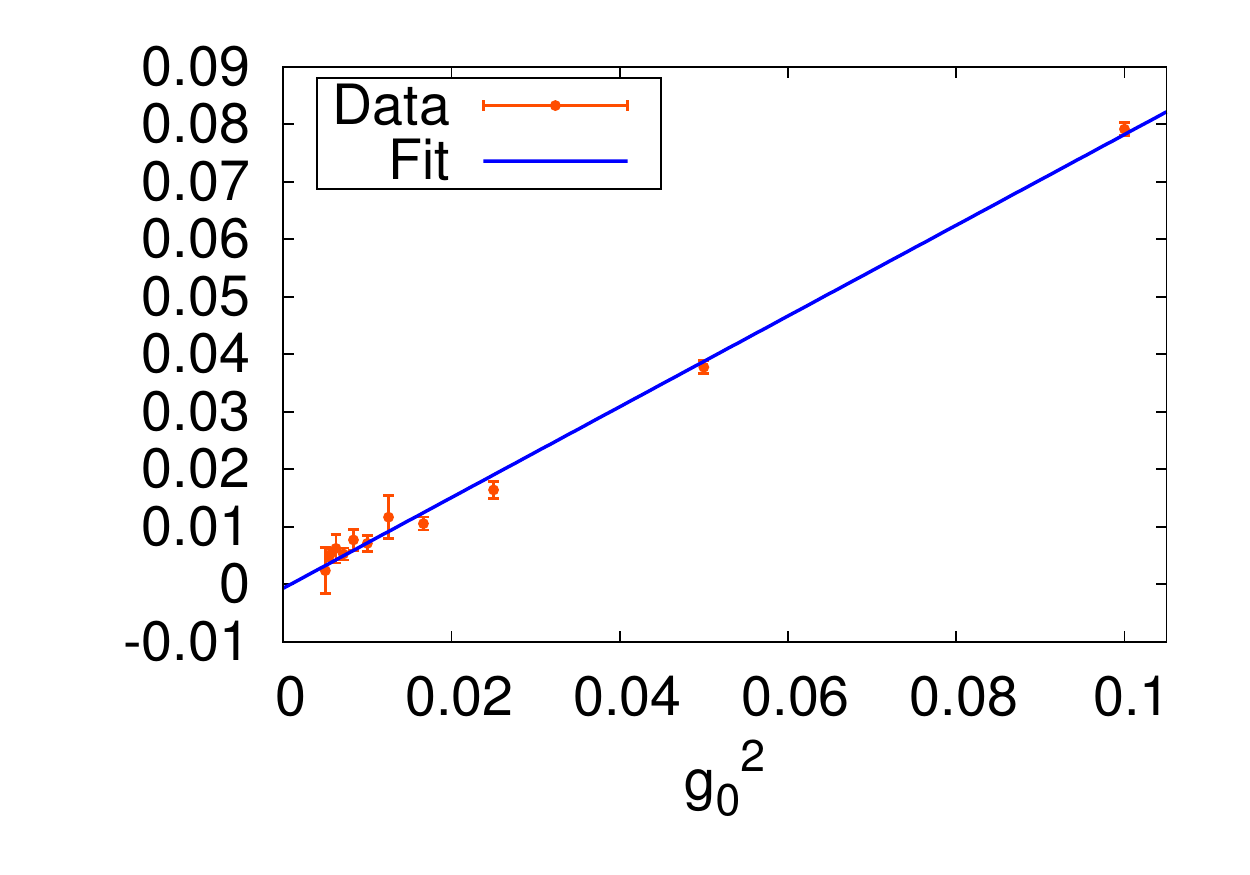}
    \caption{$c=0.4$ and $x_0/a=4$.}
  \end{subfigure}

  \caption{Two examples of the perturbative prediction versus a large
    $\beta$ pure gauge simulation. The fit to a linear
    behavior in $g_0^2$ intercepts at zero with a precision of about
    $10^{-4}$.}  
  \label{fig:lb}
\end{figure}

Finally we have performed some simulations with the \texttt{openQCD}
code~\cite{Luscher:2012av} at small values of the bare coupling in a
pure $SU(3)$ gauge theory. Using a $8^3\times 7$ lattice and varying 
the value of the bare coupling ($\beta = 60, 120, 240,
360, 480, 600, 720, 840, 960, 1080, 1200$) we compare the analytical
lattice prediction and the numerical results after collecting 10000
measurements of the gradient flow coupling for each value of
$\beta$. We use the clover definition for $G_{\mu\nu}$ to compute the
value of 
\begin{equation}
  t^2\langle E(t,x_0)\rangle|_{t=c^2L^2/8}  \,.
\end{equation}
The lattice computation of
\begin{equation}
  t^2\hat{\mathcal E}_0(t,x_0)  = 
  t^2\left[\hat{\mathcal E}_0^s(t,x_0)  +
  \hat{\mathcal E}_0^m(t,x_0)\right]
\end{equation}
can be checked in the following way: plotting 
\begin{equation}
  \frac{\langle
      E(t,x_0)\rangle - \hat{\mathcal E}_0(t,x_0)} 
    {\hat{\mathcal E}_0(t,x_0)}\Big|_{t=c^2L^2/8} = \mathcal
    O(g_0^2)  
\end{equation}
versus $g_0^2$ one expects a linear behavior with zero intercept
for all values of $c$ and $x_0$. A couple of typical cases are shown in
figure~\ref{fig:lb}, while table~\ref{tab:lb} shows the results of
the $\chi^2/{\rm dof}$ of the fits and the intercepts for all values
of $c$ and $x_0$.

\begin{table}
  \centering
  \begin{tabular}{c|lllllll}
  \bottomrule
  c & $x_0=1$ & $x_0=2$ & $x_0=3$ & $x_0=4$ & $x_0=5$ & $x_0=6$ &\\
  \hline
  \multirow{2}{*}{0.3} &
  1.77 &   1.53 &  1.31 &  1.68 &  2.10 &  1.70 & $\chi^2/{\rm ndof}$ \\
  & $5(7)$ & $-2(5)$ & $-6(5)$ & 
  $-8(5)$  & $-4(6)$ & $4(6)$ &
  Intercept$\times 10^4$ \\
  \hline
  \multirow{2}{*}{0.4} &
  1.37 &  0.98 &  0.89 &  1.35 &  1.34 & 0.91 & $\chi^2/{\rm ndof}$ \\
  & $0(1)$ & $0(7)$ & $-4(5)$ & 
  $-7(6)$  & $-3(8)$ & $5(8)$ &
  Intercept$\times 10^4$ \\
  \toprule
\end{tabular}


  \caption{Parameters of the fits to the large $\beta$
    simulations. All the fits have a good quality and the intercepts
    are zero within errors, with uncertainties of the order of
    $10^{-4}$. }
  \label{tab:lb}
\end{table}

All intercepts are of the order $10^{-4}$ and compatible with zero 
within errors. The difference in $\hat{\mathcal E}_0(t,x_0)$ for
different values of $c$, $x_0$ or between the continuum and the
lattice result varies between 5\% and 10\%.
We note that this last test is highly non-trivial since it is done
for arbitrary $x_0$ at $\rho\ne 1$ on a small lattice where cutoff
effects tend to be larger.


\section{Definition of the flow coupling}
\label{sc:coupling}

Using our continuum result 
\begin{eqnarray}  \nonumber
  \mathcal N(c,\rho, x_0/T) &=& 
  \frac{c^4(N^2-1)}{128\rho} \sum_{\mathbf n, n_0} 
  e^{-c^2\pi^2 (\mathbf n^2 + \frac{1}{4\rho^2} n_0^2)}  \\  
  &&\times  
  \frac{2\mathbf n^2s_{n_0}^2(x_0) + 
      (\mathbf n^2 + \frac{3}{4\rho^2}n_0^2)c_{n_0}^2(x_0)}{\mathbf n^2 +
      \frac{1}{4\rho^2}n_0^2} \quad 
\end{eqnarray}
we define the gradient flow coupling for non-abelian gauge theories in
the SF by means of
\begin{align}
  \overline{g}^2_{\rm GF}(L) &= 
  \left[ {\mathcal{N}}^{-1}(c,\rho,x_0/T)\cdot t^2\langle E(t,x_0)\rangle  \right]_{t={c^2L^2}/{8}} 
  \;.
\end{align}
This definition of the coupling is valid if the gauge field
is coupled to fermions in arbitrary representations. As the reader may have
noticed the scheme that defines the coupling depends not only on the
quantities $c,\rho, x_0$, but also on the value of the fermionic phase
angle and the background field. In the simulations
of the 
Schr\"odinger functional it is customary to include a phase angle
$\theta$ in the fermionic spatial boundary conditions. In
principle different values of $\theta$ are different
schemes, although we have observed in some practical situations that
the difference of the gradient flow coupling between $\theta=0$ and
$\theta=0.5$ is below the 2\%.  

Up to now we have worked
exclusively with zero background fields, but the generalization to
other values is straightforward. It only requires the modification of
the heat kernels to preserve the value of the boundary fields and
a modified form of the propagator~\cite{Weisz:int1996}. Nevertheless common
wisdom suggests that cutoff effects are reduced for zero background
field, therefore we prefer to work in this scheme. 
In this case the definition of the coupling is also symmetric
about $x_0=T/2$ and we choose that value to minimize boundary effects.
Also choosing $\rho = T/L = 1$ seems reasonable and leaves us with a
one-parameter family of couplings, parametrized by the smoothing ratio 
$c$. 

\begin{figure}[t]
  \centering
  \includegraphics[width=\textwidth]{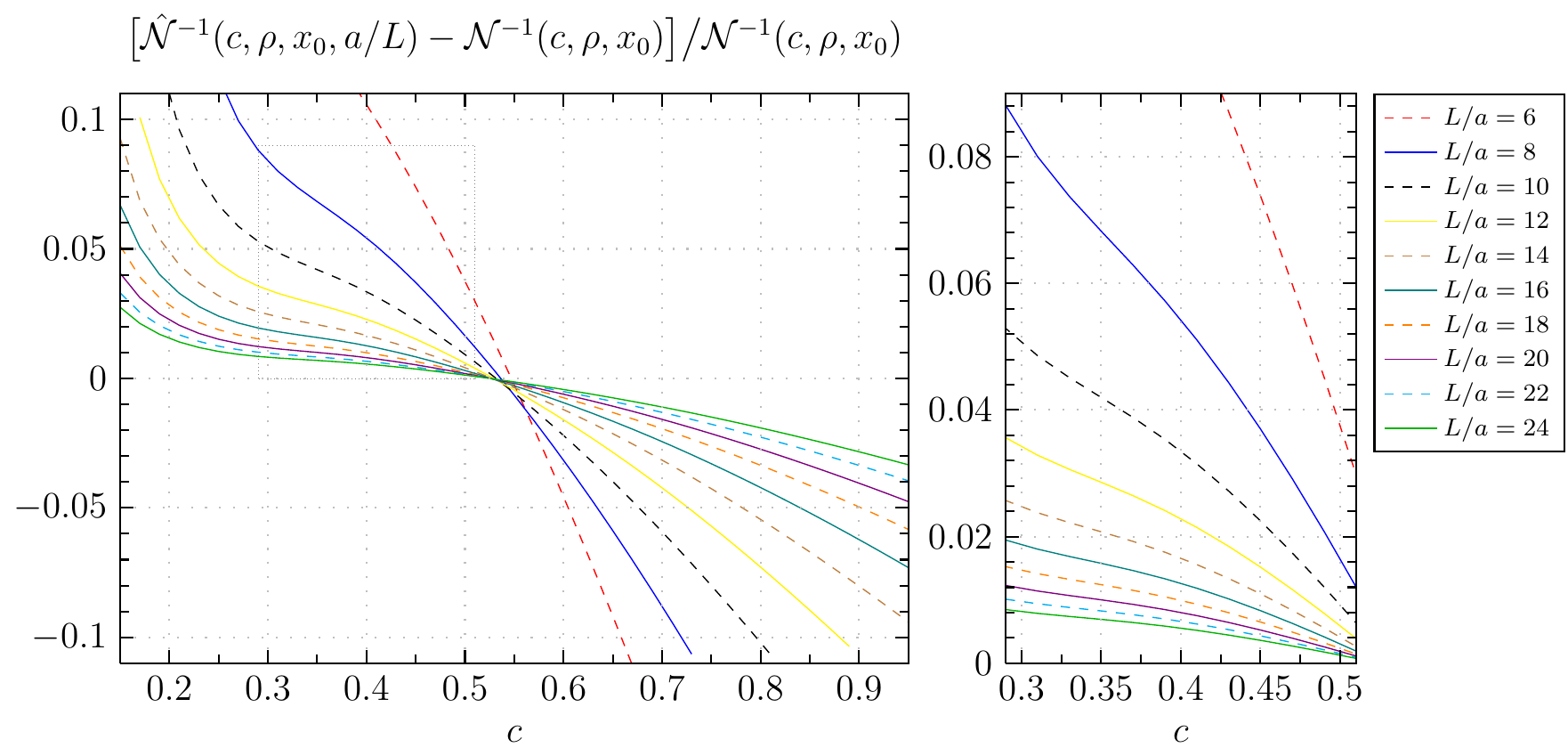}
  \caption{Leading order relative cutoff effects of the gradient flow
           coupling as a function of the smoothing ratio $c$ for 
           different $L/a$ at $\rho=1$ and $x_0=T/2$.} 
  \label{fig:cutoff}
\end{figure}

By comparing the lattice and continuum behavior of the energy density
as a function of $c$ we can compute the leading order size of cutoff
effects in the gradient flow coupling. As the reader can see in
figure~\ref{fig:cutoff}, the cutoff effects are large for small values
of $c$, reach a minimum around $c\sim 0.5$ and then grow again. We
recall that with $c=0.5$ the smoothing radius is equal to $L/2$, and
therefore one is effectively smoothing over all the lattice.
For $c=0.3$ cutoff effects are smaller than 10\% for a lattice of
size $L/a\ge8$, while for $c=0.4$ even the $L/a=6$
lattice has cutoff effects of about 10\%. 

This figure suggests using $c=0.5$ as a preferred scheme, but later,
when lattice simulations enter into the game, we will see that the
statistical errors of the coupling also grows with $c$, and therefore
in practice it is better to stay with $c\in[0.3,0.5]$, probably depending
on the particular case, but this is the subject of the next section.

We would also like to comment that if one is performing numerical
simulations with the Wilson gauge action, one can benefit from smaller
cutoff effects by using the lattice prediction to normalize the
coupling. Defining
\begin{eqnarray}
  \nonumber
  \hat{\mathcal N}(c,\rho,x_0/T, a/L) &= &
  \frac{(N^2-1)c^4}{128 \rho} \sum_{\mathbf p, p_0}
  e^{-\frac{L^2c^2}{4}(\hat{\mathbf p}^2 + \check p_0^2)}
  \left\{
    \frac{\mathring{\mathbf p}^2\cos^2(ap_i/2) - (\mathring
      p_i\cos(ap_i/2))^2}{\hat{\mathbf p}^2 +
      \check p_0^2}\hat s_{p_0}^2(x_0) \right.\\
  && + \left. \frac{\mathring{\mathbf p}^2\cos^2(ap_0/4) +
      \frac{1}{4}\hat p_0^2\cos^2(ap_i/2)
      }{\hat{\mathbf p}^2 + 
      \check p_0^2} \hat c_{p_0}^2(x_0-a/2) \right\}
\end{eqnarray}
the coupling is given by
\begin{align}
  \overline{g}^2_{\rm GF}(L) &= 
  \left[\hat{\mathcal{N}}^{-1}(c,\rho,x_0/T,a/L)\cdot t^2\langle E(t,x_0)\rangle  \right]_{t={c^2L^2}/{8}} 
  \;.
\end{align}
Obviously both definitions of the coupling differ only by cutoff effects. 

We finally want to mention that it is possible to define analogous
couplings by using only the spatial components $\langle
G_{ik}(t)G_{ik}(t)\rangle$. In a lattice 
simulation one stays further away from the boundaries by not
including plaquettes with links in the time direction. This may result
in smaller cutoff effects, although this point needs further
investigations. 


\section{Non-perturbative tests}
\label{sc:nptests}

\def\gbGF{\overline{g}_{\rm GF}}

\begin{figure}[t]
        \begin{center}
\includegraphics[width=0.8\textwidth]{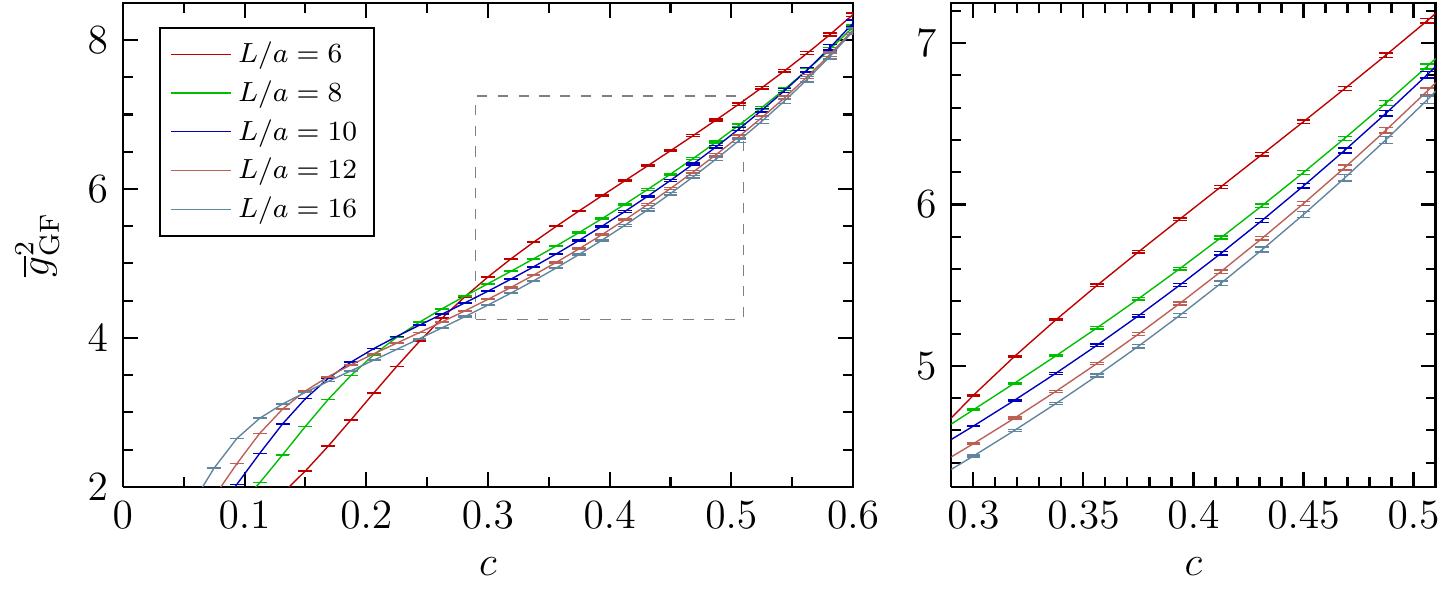}
        \end{center}
        \vskip-1.5em
        \caption{The gradient flow coupling as function of the flow time
          through $c=\sqrt{8t}/L$ for our lattices
          $L_1/a=6,8,10,12,16$ defined by a line
          of constant physics as described in the text. The right plot is a
          zoom of the dashed box in the left plot. The uncertainties are
          barely visible at this scale.}
        \label{fig:gl1}
      \end{figure}

\begin{figure}[t]
        \begin{center}
\includegraphics[width=0.8\textwidth]{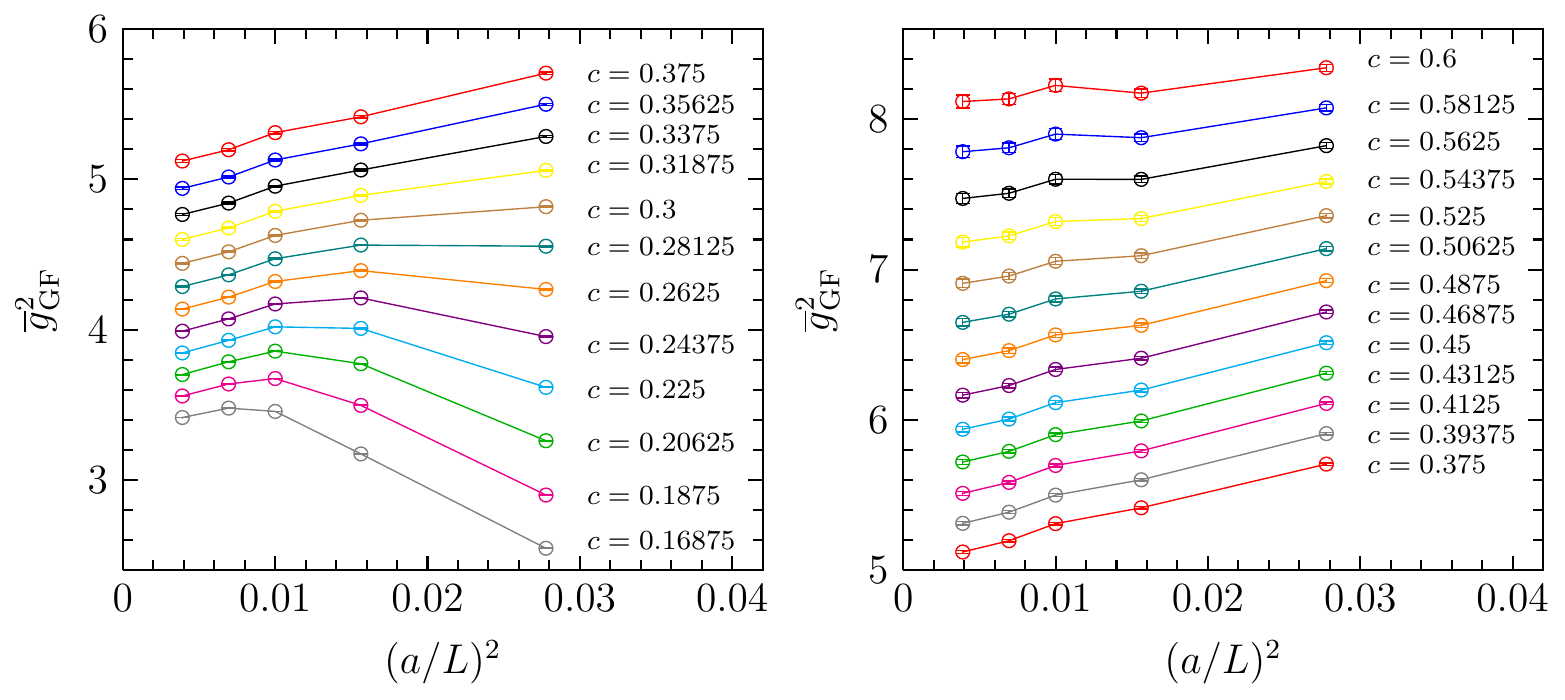}\hfill
        \end{center}
        \vskip-1.5em
        \caption{A global view on the gradient flow coupling results
          for all five ensembles and $c \ge c_{\rm min}=\max(\{a/L\})$. 
          The connecting lines are drawn to guide the eye along results 
          at constant smoothing ratio with values given in the plot.}
        \label{fig:NPglobal}
\end{figure}

In this section we would like to analyze the gradient flow coupling
numerically. We want to estimate both the size of cutoff effects and
the numerical cost of evaluating the new gradient flow coupling. The
main result 
of this section is that both quantities depend on the particular
scheme via the parameter $c$. When $c$ is increased cutoff effects
decrease, but the numerical cost increases. We find that the window of
values $c\in[0.3,0.5]$ allows a very precise determination with a
mild continuum extrapolation. 

\subsection{Line of constant physics}

As framework for our tests we choose a set of 
$N_{\rm f}=2$ Schr\"odinger functional simulations at a line of 
constant physics as given through
\begin{align}
        \overline{g}^{2}_{\rm SF}(L_1) &\equiv u = 4.484  \qquad\text{and}\qquad m(L_1) = 0 \;,
        \label{eq:lcp1}
\end{align}
where $\overline{g}^{2}_{\rm SF}$ is the traditional SF coupling 
and $m$ the renormalized PCAC mass. 
From reference~\cite{Fritzsch:2012wq} we know that the physical volume
is roughly $L_1\sim 0.4\,{\rm fm}$.
The available five different ensembles are lattices with
$L/a\in\{6,8,10,12,16\}$   
at $T=L$  with vanishing boundary gauge fields and a
fermionic  phase angle $\theta=0.5$.  Each ensemble consists
of at least 8000 configurations separated by $\tau_{\rm meas}=10$ molecular 
dynamic units (MDU). We refer the reader to the appendices
in~\cite{Blossier:2012qu} for any unexplained detail concerning the
physics and run parameters. 

We would like to measure the value of the gradient flow coupling in these
ensembles. Since they have been tuned to have constant SF
coupling, equivalent to constant volume 
in the continuum, we define the function 
\begin{align}
           \Omega(u;c,a/L) &= 
           \left[ \hat{\mathcal{N}}^{-1}(c,1,1/2,a/L)\cdot t^2\langle
             E(t,T/2)\rangle  \right]_{t=c^2L^2/8}^{u=4.484,\,m_{\rm
               sea}=0,\,\theta=0.5} 
\end{align}
that at fixed $c$ has the gradient flow coupling as continuum limit 
\begin{equation}
  \gbGF^2(L) = \omega(u;c) \equiv \lim_{a/L\rightarrow 0}
  \Omega(u;c,a/L) \,. 
\end{equation}

We will also use some ensembles with larger lattices ($L_1/a=24,32,40$)
but lower statistics. These have been defined by a slightly different 
line of constant physics: 
\begin{align}
        \overline{g}^{2}_{\rm SF}(L_1/2) &\equiv \tilde u = 2.989  \qquad\text{and}\qquad m(L_1/2) = 0 \;,
        \label{eq:lcp2}
\end{align}
that is, at a fixed SF coupling corresponding to half the scale $L_1$. 
Both LCP's are related through the step scaling function in two-flavour QCD~\cite{DellaMorte:2004bc},
\begin{align}
       \overline{g}^{2}_{\rm SF}(L_1) = \sigma( 2.989 ) = 4.484(48) \,,
        \label{}
\end{align}
and thus differ only by cutoff effects. The statistics for this second
set of ensembles is smaller ($\sim 800$
measurements). For these additional lattices we
define a function similar to $\Omega$ according to
\begin{align}
          \widetilde \Omega(u;c,a/L) &= 
           \left[ \hat{\mathcal{N}}^{-1}(c,1,1/2,a/L)\cdot t^2\langle
             E(t,T/2)\rangle  \right]_{t=c^2L^2/8}^{u=\sigma(\tilde u),\,m_{\rm
               sea}=0,\,\theta=0.5} \,,
\end{align}
that also has the same continuum limit. 

All ensembles have been tuned to have constant SF
coupling only with some statistical accuracy. This propagates into
an uncertainty in the determination of the functions $\Omega(u;c,a/L)$
and $\widetilde \Omega(u;c,a/L)$. This error
can be estimated by simple propagation of errors, for example applied
to $\Omega(u;c,a/L)$ we have
\begin{equation}
  \label{eq:prop}
  \delta \Omega(u;c,a/L) = \left|\frac{\partial \Omega}{\partial
      u}\right|\delta u \,. 
\end{equation}
To evaluate this uncertainty we use another ensemble (labeled $12^*$
in table~\ref{tab:gbGF}) with 
a slightly different value of $\beta$ but also tuned to have vanishing
quark mass. By evaluating both the SF coupling and $\Omega$ on this
ensemble we can numerically estimate the derivative in
equation~\eqref{eq:prop}%
\footnote{%
For $c=0.3, 0.4, 0.5$ we obtain $(\partial\Omega/\partial u)=0.7, 1.1, 1.5$ 
respectively.
}. 
This source of error in fact dominates the error budget of
$\Omega(u;c,a/L)$, which anticipates that the new coupling is
numerically more precise.

\begin{table}[t]
        \small
        \centering
        \renewcommand{\arraystretch}{1.0}
        \begin{tabular}{cllllllllllll}\toprule
                         $L/a$  &  $6$              &  $8$               &  $10$             &  $12$             &  $16$              &  \!\!$12$*      \!\!      \\
                       $\beta$  &  $5.2638$         &  $5.4689$          &  $5.6190$         &  $5.7580$         &  $5.9631$          &  \!\!$5.8120$   \!\!      \\
            $\kappa_{\rm sea}$  &  $0.135985$       &  $0.136700$        &  $0.136785$       &  $0.136623$       &  $0.136422$        &  \!\!$0.136617$ \!\!      \\
                $N_{\rm meas}$  &  $12160$          &  $8320$            &  $8192$           &  $8280$           &  $8460$            &  \!\!$2392$     \!\!
\\\cmidrule(lr){1-6}
 $\overline{g}^2_{\rm SF}(L_1)$ &  $4.423(75)$      &  $4.473(83)$       &  $4.49(10)$       &  $4.501(91)$      &  $4.40(10)$        &  \!\!$4.218(49)$\!\!
\\\cmidrule(lr){1-6}
            $\Omega(u;0.3,a/L)$ &  $4.818(55)$      &  $4.728(61)$       &  $4.627(73)$      &  $4.518(67)$      &  $4.441(74)$       &  \!\!--         \!\!      \\
      $\overline{g}^2_{\rm GF}$ &  $4.8178(46)$ &  $4.7278(46)$  &  $4.6269(47)$ &  $4.5176(47)$ &  $4.4410(53)$  &  \!\!$4.310(8)$ \!\!      \\
               $\tau_{\rm int}$ &  $0.57(2)$        &  $0.51(2)$         &  $0.62(3)$        &  $0.66(3)$        &  $0.92(6)$         &  \!\!$0.67(6)$  \!\!      \\
        $R_{\rm NS}\times 10^{3}$ &  $0.95(2)$        &  $0.97(2)$         &  $1.02(3)$        &  $1.04(3)$        &  $1.20(4)$         &  \!\!--         \!\!
\\\cmidrule(lr){1-6}
            $\Omega(u;0.4,a/L)$ &  $6.009(80)$      &  $5.699(88)$       &  $5.60(11)$       &  $5.484(96)$      &  $5.41(11)$        &  \!\!--         \!\!      \\
      $\overline{g}^2_{\rm GF}$ &  $6.0090(86)$ &  $5.6985(86)$  &  $5.5976(97)$ &  $5.4837(97)$ &  $5.410(12)$   &  \!\!$5.182(16)$\!\!      \\
               $\tau_{\rm int}$ &  $0.55(2)$        &  $0.52(2)$         &  $0.70(4)$        &  $0.76(4)$        &  $1.24(9)$         &  \!\!$0.73(7)$  \!\!      \\
        $R_{\rm NS}\times 10^{3}$ &  $1.43(3)$        &  $1.51(4)$         &  $1.73(5)$        &  $1.77(5)$        &  $2.23(9)$         &  \!\!--         \!\!
\\\cmidrule(lr){1-6}
            $\Omega(u;0.5,a/L)$ &  $7.11(12)$       &  $6.82(13)$        &  $6.76(15)$       &  $6.66(14)$       &  $6.60(15)$        &  \!\!--         \!\!      \\
      $\overline{g}^2_{\rm GF}$ &  $7.106(14)$  &  $6.817(15)$   &  $6.761(19)$  &  $6.658(19)$  &  $6.602(24)$   &  \!\!$6.223(29)$\!\!      \\
               $\tau_{\rm int}$ &  $0.54(2)$        &  $0.57(2)$         &  $0.82(5)$        &  $0.89(5)$        &  $1.49(12)$        &  \!\!$0.82(9)$  \!\!      \\
        $R_{\rm NS}\times 10^{3}$ &  $1.97(4)$        &  $2.26(5)$         &  $2.85(9)$        &  $2.81(9)$        &  $3.6(2)$        &  \!\!--         \!\!      \\
           \bottomrule
        \end{tabular}
        \caption{Lattice run parameters (see~\cite{Blossier:2012qu}
                 for more details) and results for the gradient flow 
                 coupling at smoothing ratio $c\in\{0.3,0.4,0.5\}$.
                 Errors are computed using the $\Gamma$-method~\cite{Wolff:2003sm}.
                 We show the values of the gradient flow coupling
                 $\overline{g}^2_{\rm GF}(L_1)$ and of the function
                 $\Omega(u;c,a/L)$.
                 Furthermore, we quote the integrated auto-correlation
                 time $\tau_{\rm int}$ of $\overline{g}^2_{\rm GF}(L_1)$, 
                 estimated in units of the measurement frequency
                 $\tau_{\rm meas}=10$~MDU (which is the same for each
                 lattice), and the noise-to-signal ratio $R_{\rm NS}$. 
                 The lattice labeled $12$* is used to estimate 
                 $(\partial\Omega/\partial u)$.}
        \label{tab:gbGF}
\end{table}

\begin{table}[t]
        \small
        \centering
        \renewcommand{\arraystretch}{1.0}
        \begin{tabular}{cllllllllllll}\toprule
                         $L/a$  &   $24$               &  $32$               &  $40$               \\
                       $\beta$  &   $6.2483$           &  $6.4574$           &  $6.6380$           \\
            $\kappa_{\rm sea}$  &   $0.1359104$        &  $0.1355210$        &  $0.1351923$        \\
                $N_{\rm meas}$  &   $632$              &  $800$              &  $850$              \\
             $\tau_{\rm meas}$  &   $10$~MDU           &  $10$~MDU           &  $4$~MDU            \\ \cmidrule(lr){1-4}
$\overline{g}^2_{\rm SF}(L_1/2)$&   $2.989(30)$        &  $2.989(35)$        &  $2.989(43)$        \\ \cmidrule(lr){1-4}
$\widetilde \Omega(u;0.3,a/L)$  &   $4.405(45)$        &  $4.402(53)$        &  $4.335(71)$   \\
$\overline{g}^2_{\rm GF}$       &   $4.405(28)$    &  $4.402(40)$    &  $4.335(62)$   \\
$\tau_{\rm int}/\tau_{\rm meas}$&   $2.0(5)$           &  $3.9(1.2)$         &  $12(5)$            \\ \cmidrule(lr){1-4}
$\widetilde \Omega(u;0.4,a/L)$  &   $5.39(8)$          &  $5.46(12)$         &  $5.34(16)$         \\
   $\overline{g}^2_{\rm GF}$    &   $5.39(6)$      &  $5.46(11)$     &  $5.34(15)$     \\
$\tau_{\rm int}/\tau_{\rm meas}$&   $2.5(7)$           &  $6.1(2.1)$         &  $19(9)$            \\ \cmidrule(lr){1-4}
$\widetilde \Omega(u;0.5,a/L)$  &   $6.64(14)$         &  $6.87(25)$         &  $6.67(30)$         \\
     $\overline{g}^2_{\rm GF}$  &   $6.62(12)$     &  $6.87(24)$     &  $6.67(29)$     \\
$\tau_{\rm int}/\tau_{\rm meas}$&   $2.9(9)$           &  $7.4(2.8)$         &  $21(10)$            \\
           \bottomrule
        \end{tabular}
        \caption{Same as table~\ref{tab:gbGF} but for the second line 
                 of constant physics, eq.~\eqref{eq:lcp2}.}
        \label{tab:gbGF_2}
\end{table}

\subsection{Numerical results and computing cost}

Figure~\ref{fig:gl1} shows the gradient flow coupling as a
function of $c$ for the different ensembles. For $c\in [0.3,0.5]$ we
observe a monotonic behavior of $\overline g_{\rm GF}^2$ with
$a/L$. As figure~\ref{fig:NPglobal} shows this seems to be the 
scaling region of the gradient flow coupling for lattices 
with $L/a>8$, and therefore this is the region on which we will 
focus from now on.

\begin{figure}[t]
    \centering
    \includegraphics[width=0.5\textwidth]{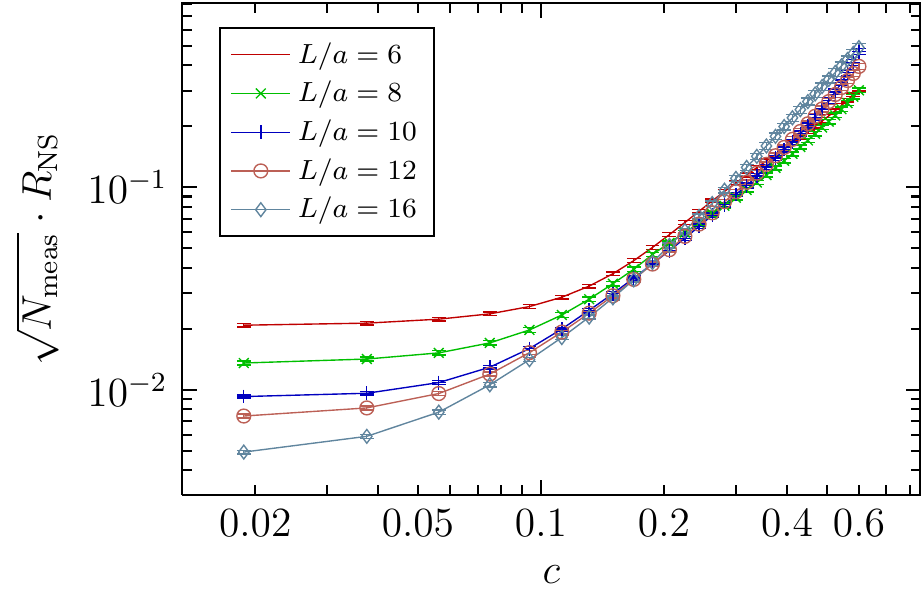}
    \vskip-0.5em
    \caption{Numerical cost of evaluating the gradient flow coupling. 
             In the plot we show $\sqrt{N_{\rm meas}}R_{\rm NS}$ as a 
             function of $c$. For the interesting values of 
             $c\ge 0.3$ there seem to be a power law scaling roughly 
             universal for all values of $L/a$.}
\label{fig:RNS}
\end{figure}

In figure~\ref{fig:RNS} we present the noise-to-signal ratio
\begin{align}
        R_{\rm NS} &= \frac{\Delta \overline{g}_{\rm GF}^2}
        {\overline{g}_{\rm GF}^2}
\end{align}
as obtained in our analysis as function of $c$ for the individual
lattices. We observe that the noise-to-signal ratio increases with
increasing $c$ (see table~\ref{tab:gbGF}). Although our statistics 
does not allow us to draw definite conclusions, we observe a
behavior compatible with a power-like scaling of $R_{\rm NS}$ with
$c$. The behavior seems to be universal and independent of
$a/L$ in contrast to the traditional SF coupling, that has a
divergent variance when approaching the
continuum~\cite{deDivitiis:1994yz}. The product  
$\sqrt{N_{\rm meas}} R_{\rm NS}$ can directly be translated in
the cost of obtaining the new gradient flow coupling with some
precision. 8000 measurements are enough to achieve a precision of
$0.1\%$ for $c=0.3$, while for $c=0.5$ the precision decreases to
$0.35\%$. 

\begin{figure}[t]
    \vskip0.25em
    \centering
    \includegraphics[width=\textwidth]{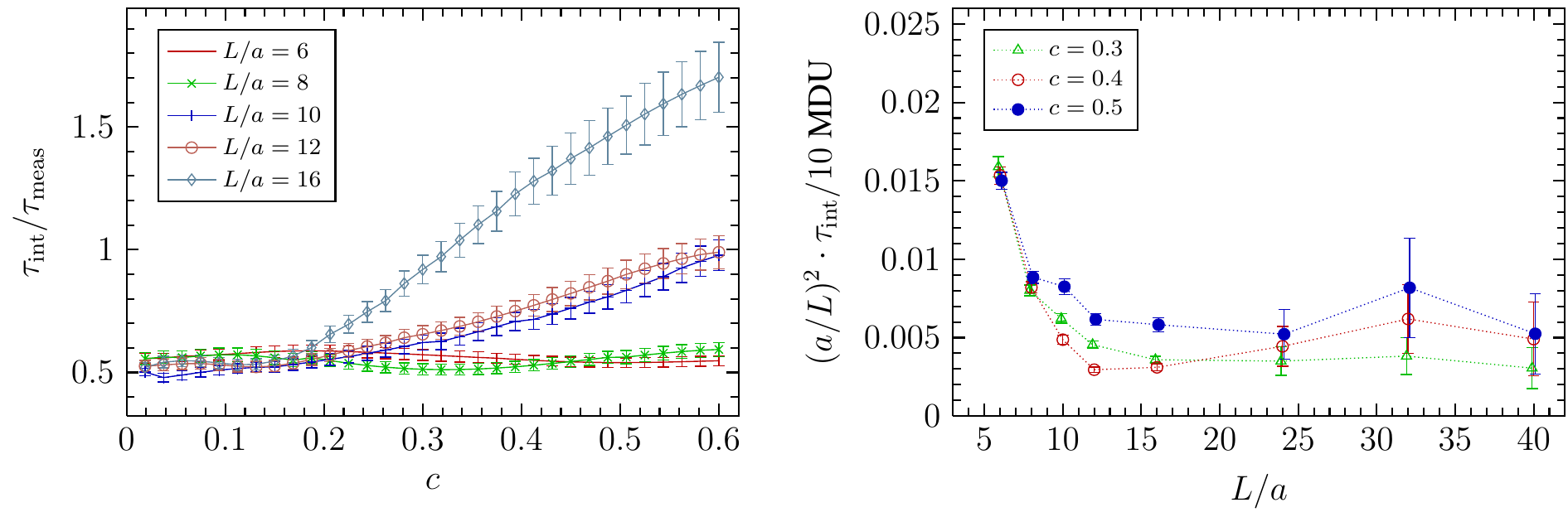}
    \vskip-0.5em
    \caption{(Left) Integrated auto-correlation time as a function 
             of $c$ for the lattices $L/a=6,8,10,12,16$. (Right) 
             $(a/L)^2\tau_{\rm int}$ as a function of $L/a$ for 
             three representative cases $c=0.3, 0.4, 0.5$ and all 
             lattices up to $L/a=40$.} 
    \label{fig:tauint}
\end{figure}

In figure~\ref{fig:tauint} we plot the integrated auto-correlation
time $\tau_{\rm int}$ for the different ensembles. The lattices 
$L/a=6,8$ have $\tau_{\rm int}\sim 0.5$ which means that the
available configurations are too far separated in Monte Carlo simulation 
time to detect any auto-correlations in the chain. The
$L/a=10,12,16$ lattices show a clear increase of auto-correlations with 
increasing $L/a$. As can be inferred from figure~\ref{fig:tauint}
this increase is compatible with a scaling $\sim 1/a^{2}$ towards the
continuum.
This is much in accordance with the conjecture 
of~\cite{Luscher:2011qa,Luscher:2011kk} 
for the scaling behaviour of the HMC algorithm in an interacting
theory, since we do not expect non-zero topological charge sectors to
contribute significantly at this small physical volume. 

\subsection{Cutoff effects}

\begin{figure}[t]
    \centering
    \includegraphics[width=\textwidth]{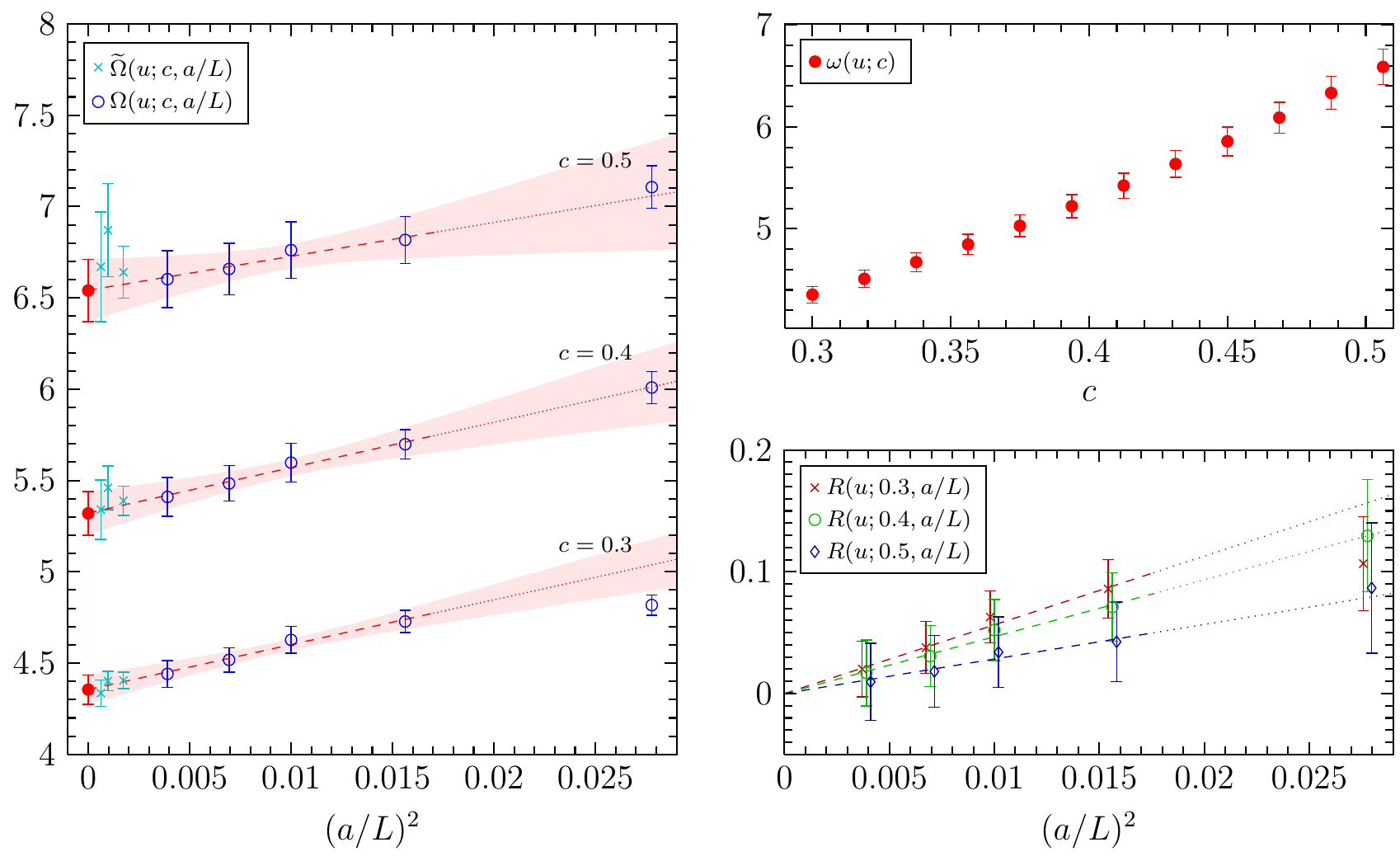}
    \vskip-0.5em
    \caption{(Left) Continuum extrapolations of $\Omega(u;c,a/L)$
      for three different values of $c = 0.3,0.4,0.5$. Only the
      lattices with $L/a=8,10,12,16$ are used for the fit, but the
      $L/a=6$ lattice as well as the larger lattices $L/a=24,32,40$
      are in the plot. (Top right) The value of the gradient flow
      coupling in the continuum as a function of $c$. (Bottom right)
      Relative size of the continuum extrapolation for the three
      representative cases $c=0.3, 0.4, 0.5$.}
\label{fig:NPtests}
\end{figure}

For the continuum approach of the function $\Omega(u;c,a/L)$ 
(and also of $\widetilde \Omega(u;c,a/L))$ we observe a behaviour 
dominated by a linear scaling in $(a/L)^2$. Hence, we choose
\begin{align}
        \Omega(u;c,a/L) &= \omega(u;c)\left\{ 1 + A(u;c)\cdot(a/L)^2
        \right\} 
\end{align}
as fit ansatz to extract the continuum limit $\omega(u;c)=\gbGF^2(L)$ and the 
leading cutoff effects $A(u;c)$. Figure~\ref{fig:NPtests} shows examples of
these extrapolations for three representative cases
$c=\{0.3,0.4,0.5\}$. We observe that cutoff effects decrease with increasing
$c$. Quantitatively this can be estimated by looking at the relative size
of cutoff effects through the ratio  
\begin{align}
        R(u;c; a/L) &= \frac{\Omega(u;c,a/L)-\omega(u;c)}{\omega(u;c)} \;, 
\end{align}
that decreases by a factor 2 when $c$ is increased from $0.3$ to $0.5$
(see figure~\ref{fig:NPtests}).

Since in general cutoff effects of the SF step scaling function are very small, 
and given the fact that cutoff effects of $\Omega(u;c,a/L)$ change with $c$,
we think that the cutoff effects in $\Omega(u;c,a/L)$ and 
$\widetilde\Omega(u;c,a/L)$ are dominated by the lattice spacing dependence 
of the new gradient flow coupling. Therefore we expect 
both functions to show roughly the same scaling behavior. Although
the points corresponding to $\widetilde\Omega(u;c,a/L)$ have not been
used for the previous continuum extrapolations, we have added the
points to the plots in figure~\ref{fig:NPtests}. The data
fits well into the expected scaling behavior.

Summarizing figure~\ref{fig:NPtests} and table~\ref{tab:gbGF} we can
say that when $c$ is increased from $c=0.3$ to $c=0.5$ the relative
cutoff effects decrease by about a factor of 2. We remember from
previous sections that this decrease of cutoff effects comes at the 
expense of an increased relative statistical error (about three
times larger when going from $c=0.3$ to $c=0.5$). 

\subsection{Mass dependence}

Last but not least we have studied the mass dependence of the gradient
flow coupling  
$\overline{g}_{\rm GF}^2$ as it was done for $\overline{g}_{\rm SF}^2$ 
in~\cite{Blossier:2012qu}. For this purpose we have generated an
ensemble with $L/a=8$ at the same value of the bare coupling as the 
available one, but with a non-zero quark mass. Actually the bare parameters 
of the simulation correspond to the lattice labeled as $8^*$
in~\cite{Blossier:2012qu}, and the interested reader is encouraged to
consult the original work for more details. Defining the dimensionless
PCAC quark mass $z=L m$, we obtain  
\begin{align}
        \left.\frac{\partial \overline{g}_{\rm GF}^2}{\partial z}\right|_{u=4.484}
        &= \left\{ \begin{array}[]{l}%
                   0.19(7) \quad\text{for $c=0.3$}\\ 
                   0.17(9) \quad\text{for $c=0.4$}\\
                   \end{array}\right.
                   \;,
        \label{}
\end{align}
to be compared to the corresponding value of $1.4(4)$ for the Schr\"odinger 
functional coupling. The mass dependence of the gradient flow coupling as 
defined in the present paper is smaller by an order of magnitude.


\section{Conclusions}
\label{sc:conclusions}

The gradient flow can be used to define a renormalized coupling at a
scale $\mu = 1/\sqrt{8t}$. In this work we have studied the
perturbative behavior of the gradient flow in the Schr\"odinger
functional. By setting the renormalization scale proportional 
to the linear size of the SF box, $\mu = 1/\sqrt{8t}=1/cL$, we have 
defined a family of running coupling constants valid for an arbitrary 
$SU(N)$ gauge field coupled to arbitrary fermions. Since this coupling
definition does not depend on any scale but the finite volume, it can be
used for finite-size scaling purposes. 

The coupling constant can be defined for different values of the
background field in the SF. Since one expects cutoff effects to be
smaller for the case of vanishing background field this is the case 
that we have studied in more detail. From our perturbative analysis we
have been able to study the size of  
cutoff effects to leading order. Cutoff effects tend to be relatively
large for either small or large values of $c$, but very
mild for $c\in[0.3,0.5]$. As an example for $c=0.3$ the difference
between the coupling in a $L/a=8$ lattice and the continuum is around
10\%, while for $c=0.5$ the difference between the continuum and the
value in an $L/a=6$ lattice is below 4\%. 

We have analyzed a total of five ensembles tuned to have a constant SF
coupling in a physical volume of $L\sim 0.4$~fm. The cutoff 
effects observed in the gradient flow coupling on these ensembles shows 
a similar overall behavior as expected from earlier perturbative
considerations. Provided that the smoothing ratio $c$ is choosen
wisely we can conclude that the gradient flow coupling has mild cutoff
effects even for small lattices.

We have analyzed the numerical cost of evaluating the gradient flow
coupling, and see that it increases with $c$. Nevertheless it can be
computed very precisely with a modest numerical effort. For the case
studied in detail, a box of size $L\sim 0.4$ fm, we
find that roughly 8000 measurements are enough to obtain a precision
of 0.1\% for $c=0.3$, even on our larger lattice $L/a=16$. The worst
analyzed case at, $c=0.5$, this precision drops to the still very good
figure of about 0.35\%. 

What is the most convenient scheme? From a practical point of view
there is no need to settle this discussion -- and thus the unique
definition of the scheme -- immediately since one measures the
gradient flow coupling at different values of $c$ in any case while
integrating the Wilson flow. 

In summary we conclude that the gradient flow coupling in the SF has
several practical advantages. It is valid for arbitrary $SU(N)$ gauge
fields coupled to fermions in any representation, having the
universal two loop beta function. It has mild cutoff effects and can be
cheaply evaluated numerically with high precision. It
has a very small dependence on quark masses and is naturally defined
with vanishing background field, avoiding the generation of new
ensembles just to compute the running coupling. 

We think that the computation of the next order in the perturbative
behavior of the gradient flow coupling in the SF is very interesting, 
in particular in connection with the determination of the Lambda parameter. 
This computation can also shed some light on an optimal value of $c$ for
the matching between different schemes. There are many applications
for this new coupling, but we are particularly interested in using it 
to compute the step scaling function of QCD. We believe that this can 
be achieved with a very high accuracy with a modest computational 
effort. 


\section*{Acknowledgments}

In particular, the authors would like to thank R.~Sommer for suggesting us (with
immense patience) to look into this very interesting problem and also
for guiding us in the steps of this work. 

We have profited from the generosity of many individuals who have
provided us with code for several purposes. We thank M.~L\"uscher for
making available a code to integrate the flow equations numerically,
to H.~Simma for the invaluable help in adapting the code to our needs,
and finally to M.~Marinkovic ({\tt SF-MP-HMC}) and the \texttt{openQCD}
team (M.~L\"uscher, S.~Schaefer and J.~Bulava for implementing SF 
boundary conditions) for their very useful codes that we have used at 
different stages of this project to generate new ensembles.

A.~R. also wants to thank A.~Gonzalez-Arroyo, S.~Sint, U.~Wolff and
F.~Virotta for the many illuminating discussions concerning the role
of the boundary conditions, the SF and the Wilson flow. Thanks, guys!

This work is supported by the Deutsche Forschungsgemeinschaft
in the SFB/TR~09 and the John von Neumann institute for computing. Our
numerical computations were performed on the PAX cluster at DESY, Zeuthen.

\appendix

\section{Notation}
\label{ap:not}

Momenta are always defined to be
\begin{equation}
  p_i = \frac{2\pi n_i}{L}
\end{equation}
for the periodic, spatial directions and 
\begin{equation}
  p_0 = \frac{2\pi n_0}{T}
\end{equation}
in the time direction ($x_0$). In the continuum sum over momenta are
abbreviated by
\begin{equation}
  \sum_p = \sum_{n=-\infty}^{\infty}\,,\; \text{ with } p=\frac{2\pi n}{L}
  \text{ or } p=\frac{2\pi n}{T}
\end{equation}
while in the lattice we have finite sums
\begin{eqnarray}
  \sum_{p_i} &=& \sum_{n_i=-L/2a}^{L/2a-1}\,,\; \text{ with } p_i=\frac{2\pi n_i}{L} \,,\\
  \sum_{p_0} &=& \sum_{n_0=-T/a}^{T/a-1} \,,\;\text{ with } p_0=\frac{2\pi n_0}{T}   \,.
\end{eqnarray}
If we impose Dirichlet/Neumann boundary conditions on the interval
$x_0\in[0,T]$
\begin{eqnarray}
  f(x_0)|_{x_0=0,T} &=& 0\qquad \text{Dirichlet,} \\
  \partial_{x_0}f(x_0)|_{x_0=0,T} &=& 0\qquad \text{Neumann,}
\end{eqnarray}
 the Laplacian has eigenfunctions and eigenvalues given by
\begin{eqnarray}
  s_{p_0}(x_0) &=& \sin(p_0x_0/2) = \sin\left(\frac{\pi n_0x_0}{T}\right) \,,\\
  c_{p_0}(x_0) &=& \cos(p_0x_0/2) = \cos\left(\frac{\pi n_0x_0}{T}\right) \,,
\end{eqnarray}
respectively. The corresponding eigenvalues are given through
\begin{eqnarray}
  \partial^2_{x_0} s_{p_0}(x_0) &=& -\left(\frac{p_0}{2}\right)^2 s_{p_0}(x_0) \,,\\
  \partial^2_{x_0} c_{p_0}(x_0) &=& -\left(\frac{p_0}{2}\right)^2 c_{p_0}(x_0) \,,
\end{eqnarray}
and these functions obey the completeness relations 
\begin{eqnarray}
  \frac{1}{2T}\sum_p s_p(x)s_p(x') &=& \delta_{xx'}   \,,\\
  \frac{1}{2T}\sum_p c_p(x)c_p(x') &=& \delta_{xx'}   \,,\\
  \frac{1}{T}\int_0^T {\rm d}x\, s_p(x)s_q(x) &=& \frac{1}{2}
  \left(\delta_{pq} - \delta_{p,-q} \right)  \,,\\
  \frac{1}{T}\int_0^T {\rm d}x\, c_p(x)c_q(x) &=& \frac{1}{2}
  \left(\delta_{pq} + \delta_{p,-q} \right) \,.
\end{eqnarray}
On the lattice derivatives are substituted by finite differences 
\begin{eqnarray}
  \hat \partial_xf(x) &=& \frac{1}{a}\left(f(x+a)-f(x)\right)   \,,\\
  \hat \partial_x^*f(x) &=& \frac{1}{a}\left(f(x)-f(x-a)\right) \,,
\end{eqnarray}
and defining the family of lattice momenta
\begin{eqnarray}
       \hat p_\mu = \frac{2}{a}\sin\left(a\frac{p_\mu}{2}\right)  \;,\qquad
  \mathring p_\mu = \frac{1}{a}\sin\left(ap_\mu\right)            \;,\qquad
     \check p_\mu = \frac{2}{a}\sin\left(a\frac{p_\mu}{4}\right)  \;,
\end{eqnarray}
the discrete Laplacian $\hat\partial_x\hat\partial_x^*$ has
eigenvalues/eigenfunctions  with periodic boundary conditions
\begin{equation}
  \hat\partial\hat\partial^* e^{\imath px}= -\hat p^2 e^{\imath px}\,,
\end{equation}
where $p=2\pi n/L $. With Dirichlet/Neumann boundary conditions one has
\begin{eqnarray}
  \hat\partial\hat\partial^* \hat s_p(x) &=& -\check p^2 \hat s_p(x)\qquad
  \text{Dirichlet}  \,,\\ 
  \hat\partial\hat\partial^* \hat c_p(x) &=& -\check p^2 \hat c_p(x)\qquad
  \text{Neumann} \,,
\end{eqnarray}
where now 
\begin{eqnarray}
  \hat s_p(x) &=& \sin\left(\frac{px}{2}\right)        \,,\\
  \hat c_p(x) &=& \cos\left[\frac{p(x+a/2)}{2}\right]  \,,
\end{eqnarray}
satisfy similar completeness relations
\begin{eqnarray}
  \frac{1}{2T}\sum_p \hat s_p(x)\hat s_p(x') &=& \delta_{xx'}           \,,\\
  \frac{1}{T}\sum_{x=0}^{T-1} \hat s_p(x)\hat s_q(x) &=& \frac{1}{2}
  \left(\delta_{pq} - \delta_{p,-q} \right)                             \,,\\
  \frac{1}{2T}\sum_p \hat c_p(x)\hat c_p(x') &=& \delta_{xx'}           \,,\\
  \frac{1}{T}\sum_{x=0}^{T-1} \hat c_p(x)\hat c_q(x) &=& \frac{1}{2}
  \left(\delta_{pq} + \delta_{p,-q} \right)                             \,.
\end{eqnarray}


\section{Heat kernels}
\label{ap:heat}

Here we will review some known properties of heat kernels. The
interested reader will like to read appendix C of~\cite{Sint:1995rb},
with more information of heat kernels and the SF.

\subsection{Continuum heat kernels}

Heat kernels are fundamental solutions (i.e. a solution that is a
delta source at $t=0$) to the heat equation, that in one dimension
reads 
\begin{equation}
  \frac{\partial}{\partial t}f(x,t) = \frac{\partial^2}{\partial x^2} f(x,t) \;.
\end{equation}
To obtain a solution to a given problem one has to choose appropriate 
boundary conditions. In the case of Euclidean space the heat kernel is
well known to be 
\begin{equation}
  K(x,x',t) = (4\pi t)^{-1/2}\exp\left[-\frac{(x-x')^2}{4t}\right] \,.
\end{equation}

Since the heat equation is a linear equation one can construct heat
kernels with different boundary conditions with a linear combinations
of the heat kernel in Euclidean space. In particular 
\begin{equation}
  K^P(x,x',t) = (4\pi t)^{-1/2}\sum_{n=-\infty}^\infty
  \exp\left[-\frac{(x-x'+nL)^2}{4t}\right]
\end{equation}
is, by construction, both a solution to the heat equation and is
periodic in $x$ with period $L$. Therefore it corresponds to the heat
kernel on $\mathbb S^1$. Following a similar reasoning it is easy 
to see that 
\begin{equation}
  K^D(x,x',t) = (4\pi t)^{-1/2}\sum_{n=-\infty}^\infty
  \left\{ \exp\left(-\frac{(x-x'+2nT)^2}{4t}\right) -
    \exp\left(-\frac{(x+x'+2nT)^2}{4t}\right) \right\}
\end{equation}
and 
\begin{equation}
  K^N(x,x',t) = (4\pi t)^{-1/2}\sum_{n=-\infty}^\infty
  \left\{ \exp\left(-\frac{(x-x'+2nT)^2}{4t}\right) +
    \exp\left(-\frac{(x+x'+2nT)^2}{4t}\right) \right\}
\end{equation}
are heat kernels with Dirichlet and Neumann boundary conditions
on the interval $[0,T]$ respectively. 
It is worth mentioning that, via the Jacobi imaginary transformation,
the periodic heat kernel is nothing more than the third Jacobi theta
function\footnote{A standard reference with the same conventions used
  here is~\cite{ww:analysis}}
\begin{eqnarray}
  K^P(x,x',t) &=& \frac{1}{L}\vartheta_3\left(\frac{\pi}{L}(x-x')|\imath
    \frac{4\pi}{L^2}t\right) = 
  \frac{1}{L} \sum_{n=-\infty}^{\infty} e^{-\left(\frac{2\pi
        n}{L}\right)^2 t}e^{\frac{2\imath\pi n}{L} (x-x')} \,. 
\end{eqnarray}
This last expression, that can also be obtained using Poisson summation
formula, turns out to be convenient for our computations.

\subsection{Discrete heat kernels}

When performing computations of the Wilson flow on the lattice we find
a type of heat equation in which the time variable is continuous but
the Laplacian is substituted by a discrete version
\begin{equation}
  \partial_t f(x,t) = \hat \partial_x\hat\partial_x^* f(x,t) \,.
\end{equation} 
Taking appropriate boundary conditions into account, we call  
fundamental solutions of this equation discrete heat kernels. 
The most easy way to construct them is by noting that one can 
formally write 
\begin{equation}
  \hat K = \exp\left\{t\hat\partial_x\hat\partial_x^*\right\}
\end{equation}
and now the task of finding the heat kernels 
is reduced to finding eigenvalues/eigenfunctions of the
discrete Laplacian with the correct boundary conditions. By recalling
the notions of appendix~\ref{ap:not} one can immediately write
\begin{eqnarray}
  \hat K^P(x,x',t) &=& \frac{1}{L}\sum_p e^{-\hat p^2 t}e^{\imath p(x-x')}        \,,\\
  \hat K^D(x,x',t) &=& \frac{1}{T}\sum_p e^{-\check p^2 t}\hat s_p(x)\hat s_p(x') \,,\\
  \hat K^N(x,x',t) &=& \frac{1}{T}\sum_p e^{-\check p^2 t}\hat c_p(x)\hat c_p(x') \,.
\end{eqnarray}

\subsection{Properties}

The following properties are straightforward, but fundamental to
easily reproduce our results:
\begin{eqnarray}
  \int_0^L \!\!{\rm d}x'\, K^P(x,x',t) e^{\imath px'} &=& e^{-p^2t}e^{\imath px}  \,,\\
  \int_0^T \!\!{\rm d}x'\, K^D(x,x',t) s_p(x') &=&
  e^{-\left(\frac{p}{2}\right)^2t}s_p(x)      \,,\\ 
  \int_0^T \!\!{\rm d}x'\, \frac{\partial K^D(x,x',t)}{\partial x} c_p(x') &=&
  \imath pe^{-\left(\frac{p}{2}\right)^2t}c_p(x)  \,,\\ 
  \int_0^T \!\!{\rm d}x'\, K^N(x,x',t) c_p(x') &=&
  e^{-\left(\frac{p}{2}\right)^2t}c_p(x) \,.
\end{eqnarray}
These relations have discrete analoga
\begin{eqnarray}
  a\sum_{x'=0}^{L-a} \hat K^P(x,x',t) e^{\imath px'} &=& e^{-\hat
    p^2t}e^{\imath px}  \,,\\ 
  a\sum_{x'=0}^T \hat K^D(x,x',t) \hat s_p(x') &=&
  e^{-\check p^2t}\hat s_p(x) \,,\\ 
  a\sum_{x'=0}^T \frac{\partial \hat K^D(x,x',t)}{\partial x} \hat c_p(x') &=&
  \imath \mathring p \cos(ap/2)e^{-\check p^2t}\hat c_p(x)  \,,\\ 
  a\sum_{x'=0}^T \hat K^N(x,x',t) \hat c_p(x') &=&
  e^{-\check p^2t}\hat c_p(x) \,.
\end{eqnarray}
Finally we note that the construction of heat kernels in more than one
dimension is done simply by multiplying one-dimensional heat
kernels. For example 
\begin{equation}
  \hat K_{\rm SF}(x, x', t) = \left[\prod_{i=1}^3 \hat K^P(x_i,x_i',t)
  \right]  \hat K^D(x_0, x_0', t)
\end{equation}
is a 4-dimensional  heat kernel with periodic b.c. in the three space 
dimensions and Dirichlet b.c. in the time dimension.



\section{Gauge fixing and gluon propagator in the SF}
\label{ap:prop}

\subsection{Gauge fixing and boundary conditions}

Here we will deal with gauge fixing in the lattice formulation of the
SF. Basically we adapt the contents of section 6 of~\cite{Luscher:1992an} 
to our specific problem of zero background field. 
In the SF admissible gauge transformations must leave the boundary
conditions of the gauge field invariant. In our particular case of
zero background field this means that 
\begin{equation}
  U_i(x)|_{x_0=0,T} = 1 \;,\qquad  i=1,2,3\;,
\end{equation}
implying that the gauge functions have to be spatially constant at
$x_0=0, T$. Moreover 
functions that are constant \emph{everywhere} do not change the vacuum 
configuration $U_\mu(x)=1$ at all. Therefore the gauge directions that
have to be fixed can be identified with $SU(N)$ valued functions
that are constant at $x_0=0$ and equal to 1 at $x_0=T$. The Lie
algebra of this group of transformations consists of $\mathfrak{su}(N)$ 
valued functions $w(x)$ with boundary conditions
\begin{eqnarray}
  w(x)|_{x_0=0} &=& {\rm constant}\,, \\
  w(x)|_{x_0=T} &=& 0 \,.
\end{eqnarray}
Accordingly, the quantum fluctuations of the gauge fields are $\mathfrak{su}(N)$ 
valued functions, with the time component defined for $0\le x_0 < T$ and 
the spatial ones for $0\le x_0 \le T$. The boundary conditions for these 
fields are
\begin{equation}
  A_k(x)|_{x_0=0,T} = 0 \,.
\end{equation}
One can define a scalar product in these Lie algebras
\begin{eqnarray}
  (w, u) &=& a^4\sum_{x,a} w^a(x)u^a(x) \;,\\
  (A_\mu, B_\mu) &=&  a^4 \sum_{x,a,\mu} A_\mu^a(x)B_\mu^a(x) \;.
\end{eqnarray}
The lattice forward derivative $\hat \partial_\mu$ maps any
infinitesimal gauge transformation $w(x)$ to the gauge field
$\hat\partial_\mu w(x)$. As the reader can check the boundary
conditions for $w(x)$ imply the correct boundary conditions for the
gauge field $\hat\partial_\mu w(x)$. The corresponding operator
is denoted by $d$. Its adjoint maps any gauge field into an 
infinitesimal gauge transformation and is defined by
\begin{equation}
  (d^*A, w) = - (A, dw) \,.
\end{equation}
The explicit transformation done by $d^*$ is
\begin{equation}
  d^* A(x) = \left\{
    \begin{array}{ll}
            \hat\partial^*_\mu A_\mu(x)           & \text{ if } 0<x_0<T \,,\\
            \frac{a^2}{L^3}\sum_{\mathbf x}A_0(x) & \text{ if } x_0=0 \,,\\
            0                                     & \text{ if } x_0=T \,.\\
    \end{array}
  \right.
\end{equation}
This is precisely the gauge fixing function that we have to add to 
the action. In explicit form it is given by
\begin{equation}
  S_{\rm gf} = \frac{a^4}{2\xi}\sum_{\mathbf
      x,0<x_0<T,b}\left[\hat \partial^*_\mu A_\mu^b({\bf
        x},x_0)\right]^2 +  
           \frac{a^4}{L^6} \left[ \sum_{\mathbf x, b}
             A_0^b({\bf x},0)\right]^2 \;. 
\end{equation}
To understand how the boundary conditions of $A_0(x)$ arise in the
continuum, one can extend the domain of definition of $A_0(x)$ to
$-a\le x_0 \le T$ and choose to fix the additional variables
with the condition
\begin{eqnarray}
  \label{eq:bcx0}
  \hat \partial_0^* A_0(x) = \left\{
    \begin{array}{ll}
            \frac{a^2}{L^3}\sum_{\mathbf x}A_0(x) & \text{ if } x_0=0  \,,\\
                                                0 & \text{ if } x_0=T  \,.
    \end{array}
    \right.
\end{eqnarray}
Now the gauge fixing function is simply given by
\begin{equation}
  d^*A = \hat\partial^*_\mu A_\mu(x) \,.
\end{equation}
Note that the additional variables are not dynamical, but 
completely determined by the previous condition. The trick is purely a
matter of convention, but now equation~\eqref{eq:bcx0} can be
interpreted as a boundary condition for $A_0(x)$. Our gauge fields have
therefore Neumann boundary conditions at $x_0=0,T$, except for the
zero momentum  mode that has mixed 
boundary conditions. 

\subsection{Gluon propagator}

In the SF the free gluon propagator on the lattice is defined as
\begin{equation}
  \langle\tilde A_\mu^a(\mathbf p,x_0) \tilde A_\nu^b(\mathbf q,y_0)
  \rangle  = L^3\delta_{ab}\delta_{\mathbf p,-\mathbf
  q}D_{\mu\nu}(\mathbf p,x_0,y_0) 
\end{equation}
where~\cite{Luscher:1996vw,Weisz:int1996}
\begin{eqnarray}
  D_{ik}(\mathbf p,x_0,y_0) &=& \delta_{ik}d(\mathbf p,x_0,y_0) + \hat
  p_i \hat p_k(\xi-1)b(\mathbf p,x_0,y_0)\,,\\
  D_{k0}(\mathbf p,x_0,y_0) &=& + \imath p_k(\xi-1)c(\mathbf p,x_0,y_0)  \,, \\
  D_{0k}(\mathbf p,x_0,y_0) &=& - \imath p_k(\xi-1)c(\mathbf p,y_0,x_0)   \,,\\
  D_{00}(\mathbf p,x_0,y_0) &=& n(\mathbf p,x_0,y_0) + (\xi-1)e(\mathbf p,x_0,y_0) \,.
\end{eqnarray}
The relevant functions, for $\mathbf p\neq 0$ are given by
\begin{eqnarray}
  d(\mathbf p,x_0,y_0) &=&
  \frac{1}{T}\sum_{p_0}
  \frac{\hat s_{p_0}(x_0)\hat s_{p_0}(y_0)}{\hat{\mathbf p}^2 + \check
    p_0^2}  \,,\\ 
  b(\mathbf p,x_0,y_0) &=&
  \frac{1}{T}\sum_{p_0}
  \frac{\hat s_{p_0}(x_0)\hat s_{p_0}(y_0)}{[\hat{\mathbf p}^2 + \check
    p_0^2]^2}  \,,\\ 
  c(\mathbf p,x_0,y_0) &=&
  \frac{1}{T}\sum_{p_0}
  \frac{\check p_0\hat s_{p_0}(x_0)\hat c_{p_0}(y_0)}{[\hat{\mathbf p}^2 + \check
    p_0^2]^2}  \,,\\ 
  n(\mathbf p,x_0,y_0) &=&
  \frac{1}{T}\sum_{p_0}
  \frac{\hat c_{p_0}(x_0)\hat c_{p_0}(y_0)}{\hat{\mathbf p}^2 + \check
    p_0^2}\,, \\
  e(\mathbf p,x_0,y_0) &=&
  \frac{1}{T}\sum_{p_0}
  \frac{\check p_0^2\hat c_{p_0}(x_0)\hat c_{p_0}(y_0)}
  {[\hat{\mathbf p}^2 + \check
    p_0^2]^2}  \,,
\end{eqnarray}
while for $\mathbf p=0$ the only functions that change are
\begin{equation}
  n(\mathbf p,x_0,y_0) = e(\mathbf p,x_0,y_0) = a+\min(x_0,y_0) \;.
\end{equation}

Note that in the continuum limit $D_{00}(\mathbf p, x_0,y_0)$ obeys
Neumann boundary 
conditions \emph{for} $\mathbf p\neq
0$, while the spatial zero momentum mode has mixed 
boundary conditions. 


\section{Adaptive size integrators for the Wilson flow}
\label{ap:int}

On the lattice the flow equation has the form
\begin{equation}
    a^2 \frac{{\rm d}V_\mu(x,t)}{{\rm d}t} = Z(V)V_\mu(x,t)  \,.
\end{equation}
Following the advice in~\cite{Luscher:2011bx}, we use the third order
Runge-Kutta scheme given by
\begin{eqnarray}
  W_0 &=& V_\mu(x,t)          \,, \\
  W_1 &=& \exp\left\{\frac{1}{4}Z_0\right\} W_0   \,,\\
  W_2 &=& \exp\left\{\frac{8}{9}Z_1 - \frac{17}{36}Z_0\right\} W_1 \,, \\
  V_\mu(x,t+a^2\epsilon) &=& \exp\left\{\frac{3}{4}Z_2
  - \frac{8}{9}Z_1 + \frac{17}{36}Z_0 \right\} W_2  \,,
\end{eqnarray}
where 
\begin{equation}
  Z_i = \epsilon Z(W_i) \,.
\end{equation}
We simply want to point out that with one extra computation, one can get a
second estimate of $V_\mu(x,t+a^2\epsilon)$
\begin{equation}
  V_\mu'(x,t+a^2\epsilon) = \exp\left\{-Z_0
  +2Z_1 \right\} W_0  \,.
\end{equation}
This last scheme is of second order, as the reader can easily check. 
Given two $N\times N$ complex matrices $A,B$, one can define a
\emph{distance} between them as follows
\begin{equation}
  {\rm dist}(A,B) = \frac{1}{N^2}\sqrt{\sum_{i,j}|A_{ij} - B_{ij}|^2} \,.
\end{equation}
This distance can be used to estimate the error made \emph{by the
lower order integrator}
\begin{equation}
  d = \max_{x,\mu}\left\{ {\rm dist}(V_\mu(x,t+a^2\epsilon),
    V_\mu'(x,t+a^2\epsilon)) \right\}   \,.
\end{equation}

With this information one can adjust the step size so that the error
in the integration does not exceed certain tolerance $\delta$. After each
integration step $\epsilon$ is updated according to
\begin{equation}
  \epsilon \longrightarrow \epsilon 0.95 \sqrt[3]{\frac{\delta}{d}}
\end{equation}
and if $d>\delta$ the integration step is repeated.

The reason why this scheme is efficient for the particular case of the
Wilson flow is related to its smoothing properties. Close to $t=0$ the
configuration is rough and therefore a very fine integration is
needed. But as $t$ increases the configuration is more and more
smooth, and one can have a very precise integration with a large
$\epsilon$. In practical cases it has saved us around a factor 4 in
the number of integration steps. It also has the advantage that
$\epsilon$ is tunned automatically, and one does not need to worry
about the step size, but only plug in the desired tolerance.


\bibliography{/home/alberto/docs/bib/math,/home/alberto/docs/bib/campos,/home/alberto/docs/bib/fisica,/home/alberto/docs/bib/computing}

\end{document}